\documentclass[10pt,a4paper]{article}

\usepackage[round, sort, numbers]{natbib}
\usepackage[cm]{fullpage}
\usepackage{amsmath, amsthm, amsfonts}
\usepackage{times,amssymb,latexsym}
\usepackage{graphicx}
\usepackage{color}
\usepackage{lineno}
\usepackage{setspace}
\usepackage[titletoc]{appendix}
\usepackage{authblk}
\usepackage{epstopdf}
\usepackage{hyperref}
\hypersetup{colorlinks=true, linkcolor=blue, filecolor=magenta,  urlcolor=cyan, citecolor=magenta}
\usepackage{xr}
\usepackage{cleveref}
\date{} 
\providecommand{\keywords}[1]
{  \small	
  \textbf{\textit{Keywords---}} #1
}

\title{Downward continuation of marine seismic reflection data:\\
An undervalued tool to improve velocity models}

\author[1]{C. E. Jimenez-Tejero \thanks{ejimenez@icm.csic.es}}
\author[1,2]{Cesar R. Ranero}
\author[1]{Valenti Sallares}
\author[1,3]{Claudia Gras}
\affil[1]{Barcelona Center for Subsurface Imaging, Institute of Marine Sciences, CSIC, Pg. Mar\'itim de la Barceloneta 37-49, 08003 Barcelona (Spain).}
\affil[2]{ ICREA at CSIC}
\affil[3]{ Currently at Frontwave Imaging}

\begin{document}


\maketitle

\begin{abstract}
The purpose of marine seismic experiments is to provide information of the structure and physical properties of the subsurface. The P-wave velocity distribution is the most commonly modelled property, usually by inversion of arrival times or waveform attributes. In wide-angle seismic reflection/refraction (WAS) experiments, arrival times of seismic phases identified in data recorded by Ocean Bottom Seismometers (OBS), are used to image relatively deep structures. Most WAS experiments have relatively low redundancy and produce robust velocity models of limited resolution. The shallow subsurface is also commonly studied with Multi-Channel Seismic (MCS) data recorded by towed streamers, a technique that is highly complementary to WAS. In this case, the recording of refractions as first arrivals is limited primarily by the streamer length and by features like water depth and the velocity structure and, in general, most refractions are masked by reflections and noise. However, MCS data of variable quality are available in many regions where no other data exist and previous work has shown that these data can also be used to retrieve velocity models through travel-time and full waveform inversion provided that first arrival information is properly extracted from the record sections. The most widely used tool to extract refractions as first arrivals from MCS recordings is the so-called downward continuation technique, which is designed for redatuming streamer field data to the seafloor. In this new virtual configuration, the early refractions transform to seismic phases that are becoming visible as first arrivals from nearly zero offset, facilitating its identification and use in travel-time tomography. However, there is limited literature, let alone available codes, to be used with the available MCS data sets. This work presents a user friendly open source HPC software for redatuming 2D streamer field data to the sea bottom for any seafloor relief. The main ingredient is the acoustic wave equation used backward in time, allowing first the redatuming of the receivers, and after, the redatuming of the sources. Assessment tools are provided to evaluate the information available after redatuming for specific data acquisition configurations. Also, we present a step-by-step analysis that defines the most important features that influence the quality of the virtual, redatumed recordings.
\end{abstract}


\section{Introduction}\label{sec:1}
The estimation of physical properties of the subsurface, such as the P- and S- wave velocity, density and anisotropy, has a broad range of applications in different geoscientific disciplines and industry sectors. This cross-disciplinary interest have prompted a sustained development of a suite of geophysical technologies to estimate them and to determine their spatial distribution. 

The most commonly modelled physical property is the P-wave velocity (V$_p$). It can have multiple purposes, as it may serve as a base to estimate other physical properties, to infer other attributes such as rock type, and, together with the S-wave velocity, to evaluate the porosity degree, and fluid or gas content, pressure and distribution. 

The information on the V$_p$ distribution may be needed as a low resolution map for seismic processing including stacking and time migration, or it may be needed as a high-resolution model for more sophisticated imaging applications like pre-stack depth migration. The resolution and uncertainty of any V$_p$ distribution is strongly dependent on the characteristics of the field data available and on the technology used to extract it. Many works have discussed the different technologies available for V$_p$ determination, which mainly include different forms of seismic data inversion, including both travel-time and waveform tomography approaches. All these methods exploit some characteristics of the input data, however, the field data potential is often not thoroughly analyzed nor fully exploited. 

Most controlled source methods used to determine the V$_p$ distribution use pre-defined attributes of either reflected or refracted seismic phases. Reflected phases at structural boundaries may abound in stratigraphical sequences, but they are absent or highly discontinuous in basement rocks (REF). Further, there is an intrinsic uncertainty between reflector position in the subsurface and the overlying V$_p$. In contrast, refracted waves are dominantly influenced by the seismic velocity of the media where they travel, and provide a more robust information of its V$_p$, but do not have direct information on the location and geometry of  structural boundaries. 

Seismic data recorded in hydrophone streamers or in seafloor seismometers are commonly used to study the sub-seafloor structure in marine settings. These data have source to receiver offsets limited to a few km, with modern-data offsets ranging from 6-15 km (\cite{Olsen:2020}, \cite{Audhkhasi:2019}), but there exist a substantial archive of data acquired with shorter streamers, of 4-6 km (\cite{Prada:2015}, \cite{Merino:2021}) from the late nineties and early two thousands, and 2.4-3 km for pre-90’s legacy data. These data are of interest because they often image regions where no other data set exists.

The vast majority of the marine streamer data have been used to obtain only low-resolution V$_p$ models through simplistic approaches, such as normal move out correction, which have an inherently high uncertainty \cite{Yilmaz:2001}. This is because these data were processed for imaging of impedance contrasts, and this type of regular processing and imaging does not require detailed V$_p$ information \cite{Yilmaz:2001b}. However, the limitations on the determination of the V$_p$ distribution are also due to the limited acquisition offsets that provide only reflected arrivals in most cases. Clear refraction arrivals allowing to improve V$_p$ modelling are only recorded in particular settings, typically in relatively shallow-water environments and/or in the most distal sections of the streamer \cite{Yilmaz:2001b}. 

Attempts to circumvent the shortcoming of V$_p$ model building with streamer data and producing better-constrained and more accurate V$_p$ models include the transformation, or redatuming, of the wavefield recorded by the streamer hydrophones to a virtual geometry with the acoustic source and hydrophone array downward continued to the depth and following the relief of the seafloor. This procedure allows collapsing the reflection wavefield and exposing the refracted arrivals masked in the originally seismic record as first arrivals from the closest receiver to the source. In this way, the arrival times can be picked and used to perform first arrival~\cite{Gras:2019} or joint refraction and reflection travel time tomography \cite{Gras:2021} or their waveforms can be inverted for V$_p$ determination~(\cite{Satish:2017}, \cite{Marjanovic:2017}, \cite{Harding:2016}).

Although this method has the potential to provide data for robust V$_p$ model building to be used for depth imaging or as initial model for waveform inversion, it has not been widely used to data possibly because no standard downward continuation method has been described in the literature with sufficient detail and no freely available user-friendly code has been released. Further, no systematic analysis has been presented to describe how offset distance, water depth and subsurface V$_p$ distribution affect or even prevent the recording of refracted arrivals in the streamer.

Here we present, describe and provide a dedicated in-house software and a short manual to apply DC to data with Seismic Unix (SU) format in a user friendly manner. Additionally, we include a series of tests to provide a guide to understand and predict the presence of refracted arrivals in streamer data and use it for V$_p$ model building after downward continuation (DC) of the records. The code is intended to produce the DC record that could be used for V$_p$ model building using a preferred tomographic approach and our guide gives some clues on how much of the record propagated to the seafloor can reliably be used for travel-time inversion.

\section{Materials and methods}\label{sec:2}
In this work, the two-way acoustic wave equation in 2D is solved to generate synthetic shot gathers with different acquisition geometries and it is also the main ingredient for redatuming the streamer field data from the sea surface to the seafloor. This section describes the implementation of the wave equation in detail and its use for the data redatuming procedure. Besides, we frame our work within the different redatuming techniques presented in the literature. Also, we describe the computational aspects related to the code and the data pre-processing techniques applied to improve the quality of the results.

\subsection{2D acoustic wave propagation}\label{subsec:2.1}
As the goal is propagating the acoustic wave through the water layer, the homogeneous, isotropic and acoustic approximations are applied to the wave equation. The wave propagation equation is solved using a finite difference scheme of 6th order in space and 2nd order in time. We apply a free surface at the top and convolutional perfectly matched layers (CPML) scheme~\cite{Pasalic:2010} to avoid artificial numerical reflections on the left, right and bottom boundaries. The wave-equation is defined differently outside and inside the PML layers. The acoustic wavefield propagation, {p}~=~{p}(\textbf{r},t), of a given source, $f_\text{s}={f}_\text{s}(\textbf{S},t)$, in a velocity model, ${v}={v}(\textbf{r},t)$, is defined as follows:
\begin{equation}\label{Eq:s1}
\frac{1}{{v}^2}\partial^2_{t}p = \partial^2_{k} {p_k} + f_\text{s} + f^\text{PML}_p \ ,
\end{equation}
\begin{equation}
f^\text{PML}_p = 
\begin{cases} 
\partial_k \cdot \psi_k(p) +\xi_k(p), ~ \textbf{r} \in \text{PML} \\
\\
0 , ~\textbf{r} \notin \text{PML} 
\end{cases}
\end{equation}
The additional term for the PML layer $f^\text{PML}_p$ is defined with the auxiliary variables $\psi_k(p)$ and $\xi_k(p)$, whose evolution for each component~(k=x,z)~and time step, $n$, is:
\begin{eqnarray}\label{Eq:s2}
\left[\psi_k(p) \right]^n &=&a_k \cdot \left[\psi_k(p) \right]^{n-1}+b_k \cdot \left[ {\partial_k p_k} \right]^n \ , \\
\left[\xi_k(p) \right]^n &=& a_k \cdot \left[\xi_k(p)\right]^{n-1} + b_k \cdot \left[ {\partial_k^2 p_k}+ {\partial_k \psi_k(p)} \right]^n \ , \\
\end{eqnarray}
where the parameters a$_k$ and b$_k$ are:
\begin{eqnarray}\label{Eq:s3}
a_k= e^{-(\sigma_k+\alpha_k)\cdot \Delta t} &,&~b_k= \frac{\sigma_k}{\sigma_k+\alpha_k}\cdot \left( a_k - 1\right) \ .
\end{eqnarray}
with $\Delta t$ the time step, $\sigma_k$ the absorption damping factor of the acoustic wave and $\alpha_k$ the real positive pole shifting factor \cite{Zhang:2010}. 

\begin{table*}[t!]
\caption[Components of the 2D-wave equation for the different geometries]{Components and solution of the 2D-wave equation (Eq.~\ref{Eq:s1}) for the different geometries and redatuming steps. First column: list of the different geometries, streamer, geometry~1 (Geom1) and geometry~2 (Geom2) and the virtual ones for geometry~1 (vGeom1) and~2 (vGeom2). The block letters \textbf{S} and \textbf{R} refer to sources and receivers respectively, and the up and down indexes refer to the datum geometry, i.e. depth of the sources and receivers at the surface and seafloor, respectively. Second column: velocity model, where the symbol $\oplus$ refers to the joint consideration of the water and Earth layers to describe the velocity model. Third column: source function, f$_s$. Forth column: wavefield function propagating in space and time. Fifth column: shot gather solution in the set of receivers,~$\textbf{R}$. The expression PG (SG) indicates the conversion to point gather (shot gather) and the functions $\text{p}$ and $\tilde{\text{p}}$ refer to the shot gathers and point gathers, respectively. The minus sign in the time, $-t$, refers to the backwards propagation. For more details, check sections \ref{subsec:2.1} and \ref{subsec:2.2}.}
\begin{center}
\begin{tabular}{c|ccc|c}
Geometry &Velocity: $v$ & Source: f$_\text{s}$&Wavefield: p& shot gather solution \\
&&&& at \textbf{r} = \textbf{R} \\
\hline
&& && \\
Streamer &$v_\text{water} \oplus v_\text{Earth}$ & Ricker$(\textbf{S}^{\text{up}}_\text{j},\text{t})$ & $ \text{p}_\text{streamer}(\textbf{r},\text{t})$ & $ \text{p}_\text{streamer}(\textbf{R}^{\text{up}},\text{t})$ \\
$(\textbf{S}^{\text{up}},\textbf{R}^{\text{up}})$&& && \\
&& && \\

Geometry 1 & $v_\text{water}\oplus v_\text{Earth}$&Ricker$(\textbf{S}^{\text{up}}_\text{j},\text{t})$& $ \text{p}_\text{Geom1}(\textbf{r},\text{t})$ & $ \text{p}_\text{Geom1}(\textbf{R}^{\text{down}},\text{t})$ \\
$(\textbf{S}^{\text{up}},\textbf{R}^{\text{down}})$&& && \\
&& && \\

Geometry 2 & $v_\text{water}\oplus v_\text{Earth}$&Ricker$(\textbf{S}^{\text{down}}_\text{j},\text{t})$ & $\text{p}_\text{Geom2}(\textbf{r},\text{t})$ & $ \text{p}_\text{Geom2}(\textbf{R}^{\text{down}},\text{t})$ \\
$(\textbf{S}^{\text{down}},\textbf{R}^{\text{down}})$&& && \\
&& && \\

\hline
&& & & \\
Virtual Geometry 1& $v_\text{water}$& $ \text{p}_\text{streamer}(\textbf{R}^{\text{up}},-\text{t})$ & $ \text{p}_{\text{vGeom1}}(\textbf{r},-\text{t})$ & $ \text{p}_{\text{vGeom1}}(\textbf{R}^{\text{down}},\text{t})$ \\
$(\textbf{S}^{\text{up}},\textbf{R}^{\text{down}})$&& && \\
&& && \\
Virtual Geometry 2&$v_\text{water}$& $\xrightarrow[\text{}]{\text{PG}} \tilde{\text{p}}_{\text{vGeom1}}(\textbf{S}^{\text{up}},-\text{t})$& $\tilde{\text{p}}_{\text{vGeom2}}(\textbf{r},-\text{t})$ &$\xrightarrow[\text{}]{\text{SG}}\text{p}_{\text{vGeom2}}(\textbf{R}^{\text{down}},\text{t})$ \\
$(\textbf{S}^{\text{down}},\textbf{R}^{\text{down}})$&& && \\
&& && \\

\end{tabular}
\end{center}
\label{Tab:1}

\end{table*}

Table~\ref{Tab:1} displays the different components conforming the acoustic wave equation (Eq.~\ref{Eq:s1}) and the shot gather solution depending on the geometry and redatuming step (first column). The up and down indexes refer to the datum geometry of the sources and receivers; up, to the depth at the  water surface (typically between 5-15 meters below the surface) and down, to the depth at the seafloor, respectively. The first three rows refer to the acquisition geometries: streamer geometry in which sources and receivers are located at the sea surface, geometry~2, equivalent to the Ocean Bottom Seismometer geometry (OBS), with sources at the surface and receivers at the seafloor, and geometry~2, with both sources and receivers at the seafloor. The second column indicates the velocity model used, defined by a water layer and the sub-seafloor layer ($v_\text{water}\oplus v_\text{Earth}$). The third column shows the source used to generate shot gathers in these geometries, defined as a Ricker wavelet typically located 5-15 m below the sea surface for the streamer and geometry~1 acquisitions (upper index 'up') and at the seafloor for the geometry~2 (upper index 'down'). Each source position $j$ is defined as:

\begin{equation}\label{Eq:s4}
\textbf{S}_j^{\text{up}} = (x_{S_j},z_{S_j}^{\text{up}});~\textbf{S}_j^{\text{down}} = (x_{S_j},z_{S_j}^{\text{down}}) \ ,
\end{equation}
where the index j, runs on the number of sources, $j=1,...,N_S$. The solution for a shot gather in the different geometries is shown in last column of the table, and it is obtained by collapsing the wavefield (fourth column) at each receiver position, i.e. at the sea surface for the streamer geometry ($\textbf{R}^{\text{up}}_i$) and at the seafloor for the geometries 1 and 2 ($\textbf{R}^{\text{down}}_i$), respectively:
\begin{equation}\label{Eq:s5}
\textbf{R}_i^{\text{up}} = (x_{R_i},z_{R_i}^{\text{up}});~\textbf{R}_i^{\text{down}} = (x_{R_i},z_{R_i}^{\text{down}}) \ ,
\end{equation}
where the index i, runs on the number of receivers, $i=1,...,N_R$.

\subsection{Downward Continuation}\label{subsec:2.2}
The downward continuation (DC) redatuming of the streamer data from the sea surface to the seafloor is done in two steps:
\paragraph{First step or receivers redatuming} First, the receivers are redatumed to the seafloor, giving as a result a virtual shot gather with acquisition geometry~1 (sources at the sea surface and receivers at the seafloor). To do so, Eq.\ref{Eq:s1} is used as well with the components displayed in Tab.~\ref{Tab:1} in the "Virtual Geometry~1" row. In this step, the traces recorded in the streamer are used as sources but propagated to the seafloor back in time, for each shot gather. Note that in this case the source is in fact a super-shot built with the traces in the streamer propagating all at once. The super-shot propagation is necessary so that the waveform transforms successfully into the new geometry. The source is indicated in Tab.~\ref{Tab:1} as $ \text{p}_\text{streamer}(\textbf{R}^{\text{up}},-\text{t})$, in which $-t$ refers to the backward propagation:
\begin{equation}
 -t=(t_{N_t}, t_{N_t-1}, ..., t_{1}) \ ,
\end{equation}
and $\textbf{R}$ refers to the position of all the traces, now sources, conforming the super-shot:
\begin{equation}
\textbf{R}^{\text{up}}=(\textbf{R}^{\text{up}}_1,...,\textbf{R}^{\text{up}}_{N_R} ) \ .
\end{equation}
The solution of the wave equation for the streamer shot gather virtually redatumed to the geometry~2, is $ \text{p}_{\text{vGeom1}}(\textbf{R}^{\text{down}},\text{t})$ (see Tab.~\ref{Tab:1}), now with the receivers located at the seafloor.

\paragraph{Second step or sources redatuming} The second and last step is the DC of the sources, giving as a result virtual shot gathers with acquisition geometry~2 (sources and receivers at the seafloor). The propagation is solved again but using this time Eq.\ref{Eq:s1} with the components displayed in Tab.~\ref{Tab:1} in the "Virtual Geometry 2" row. The sources redatuming is done similarly to the receivers redatuming but it is not as straightforward, as now, the sources propagating back in time are the point gathers, which are built using the virtual shotgathers obtained after the first redatuming step. In other words, once the virtual shotgathers from the first step are reorganized into a set of point gathers, each point gather will take the rol of a a multi-source in the wave-equation, and this is how the datum of the original sources is virtually redatumed to the sea floor. Once all the pointgathers are back-propagated to the seafloor, the last step is to re-organize them again in shot gathers for the final result of the DC process. For clarification, we will specify the details to understand and build the point gathers. A point gather (PG) is the collection of shots which contribute with a trace to that specific point in the model. To estimate the total number of point gathers in a model:
\begin{equation}
	\text{NumPGs} =1+ \frac{\text{SL} + (\text{NumShots} -1) \cdot \text{dshots}}{\text{dmodel}}~, \label{Eq:PG}
\end{equation}
where 'SL' is the streamer length, 'dshots' is the distance between shot gathers and 'dmodel' the spacial grid in the model. 
The maximum number of shots which can influence each point gather is limited by those located forward from this point at a maximum distance equal to the streamer length, SL. For example, 120 is the maximum number of shots for a point gather in the case of having SL~=~6~km and 50~meters of distance between sources. The shots located further from the SL distance in the direction of the boat, or those behind the point gather, do not contribute with any trace because the boat with the streamer is already out of radar or it has not passed by that point gather yet, respectively. Notice also that the number of shots for a point gather depends on the location of this point along the model. Thus, the point gathers at the edges of the model, beginning and end, are influenced by less shots compared to the point gathers in the middle of the model. To understand this statement, we mention an example with the first and last point gathers in the model. The first point gather, which coincides with the last receiver in the streamer (furthest from the source) of the first shot, is only influenced by that one shot. In the same way, the last point gather in the model, which coincides with the first receiver in the streamer (closest to the source) of the last shot, is only influenced by that last shot.
Once the point gathers are built, they are propagated backward in time (-$t$), from the sea surface to the seafloor. Similarly to the redatuming of receivers in the first step, the propagation of each point gather needs to be done as a super-shot so that the waveform transform correctly into the geometry~2 (see Tab.\ref{Tab:1}). Each point gather propagating as a super-shot backward in time is defined as $\tilde{\text{p}}_{\text{vGeom1}}(\textbf{S}^{\text{up}},-\text{t})$, where $\textbf{S}$ refers to the position of all the shots in the sea surface, conforming each point gather:
\begin{equation}
\textbf{S}^{\text{up}}=(\textbf{S}^{\text{up}}_1,...,\textbf{S}^{\text{up}}_{N_{S}^{PG}} ) \,
\end{equation}
being $N_{S}^{PG}$ the number of shots conforming an specific point gather.  Finally, the solution of the wave equation for the streamer shot gather redatumed to the geometry~2, is defined as $ \text{p}_{\text{vGeom2}}(\textbf{R}^{\text{down}},\text{t})$ (see Tab.~\ref{Tab:1}).

\subsection{Differences with other methodologies}
The idea of redatuming was first introduced by \cite{Berryhill:1979} and \cite{Berryhill:1984} as a wavefield extrapolation for migration (\cite{Claerbout:1976}, \cite{Claerbout:1985}, \cite{McMechan:1983}). The datuming can be considered an ingredient of migration, when migration is done as a downward continuation process (\cite{Berkhout:1981}, \cite{Barison:2011}). That is because migration intents to remove the effects of wavefield propagation by means of inverse extrapolation \cite{Berkhout:1981}. But migration, in addition to downward continuation, requires imaging principle \cite{Yilmaz:2001}.

It is difficult to compare contributions for a specific redatuming method with other ones based on the literature. There are many factors defined differently in each work depending on the extrapolation method used, such as the type of operator or approach, assumptions, the extrapolation domain, and approaches to the wave propagation, which all play important roles in the result.

In general, redatuming methods apply time shifts and amplitude factors to simulate data acquired at a different surface. First redatuming approaches had model-based operators (\cite{Berryhill:1979}, \cite{Berryhill:1984}, \cite{Shtivelman:1988}, \cite{Bevc:1997}, \cite{Mo:1997}) that require the true velocity model between datums to work well. Besides model-driven redatuming operators, there is another way to derive these operators by using data-driven approaches. In this case, the operators are extracted from the input data set using the common-focus-point (CFP) described by \cite{Berkhout:1997a} and \cite{Berkhout:1997b}). For the CFP technology, the traveltimes from the surface to a designed subsurface point are estimated from the shot gathers by focusing seismic sources iteratively toward a subsurface point while updating the involved traveltimes according to some criteria (\cite{Mulder:2005}, \cite{Tegtmeier:2008}). Three different techniques to compute the wavefield extrapolation are described in \cite{Berkhout:1981}, which are the Kirchhoff-summation approach (\cite{Berryhill:1979}, \cite{Berryhill:1984}, \cite{Shtivelman:1988}, \cite{Bevc:1997}, \cite{Tegtmeier:2008}), the plane-wave method (\cite{Gazdag:1978}, \cite{Arnulf:2011}, \cite{Cho:2016}), and the finite difference technique (\cite{Mo:1997}, \cite{Gras:2019}). All of them are wave equation datuming (WED) methods as they use extrapolation operators that account for different forms of wave propagation between the acquisition surface and the new datum level by means of the wave equation. These operators represent Green's functions. Some of them are based on simplifying assumptions such as one-way wave propagation and primary-only data (\cite{Berryhill:1979}, \cite{Berryhill:1984}, \cite{Shtivelman:1988}, \cite{Bevc:1997}, \cite{Berkhout:1997a}, \cite{Berkhout:1997b}, \cite{Tegtmeier:2008}, \cite{Harding:2016}). In contrast, two-way wave propagation redatuming operators (\cite{McMechan:1982}, \cite{McMechan:1983}, \cite{Mo:1997}, \cite{Mulder:2005}, \cite{Gras:2019}) also include multiples and refractions. Many implementations have also problems dealing with lateral V$_p$ variations due to the approximate solution to the wave equation (\cite{Berryhill:1979}, \cite{Berryhill:1984}, \cite{Shtivelman:1988}, \cite{Arnulf:2011}, \cite{Harding:2016}, \cite{Cho:2016}). More recently, cross-correlation redatuming methods have also emerged and been developed \cite{Schuster:2006} to overcome model-based limitations of wave-equation statics, or the need to specify certain events in CFP technology, among others. The cross-correlation redatuming require weighted correlation of the traces with one other, followed by summation over all sources and sometimes receivers (\cite{Schuster:2006}, \cite{Wapenaar:2014}, \cite{Wapenaar:2017}). 
 
Here, redatuming is formulated by solving the two-way full acoustic wave equation by the finite difference approach in the space-time domain. This scheme has a simple implementation, includes all data with no prior phase selection, can easily be adapted to irregular new datum surfaces being stable to steeply dipping areas, and allows to deal with arbitrary V$_p$ models. Its drawbacks are the ones associated to the finite difference scheme, mainly the numerical dispersion and, since it is a model-based redatuming approach, the need of having a sufficiently good V$_p$ model to construct the operators. But the accuracy of V$_p$ becomes a minor problem in the case of applying it through the water column.

\subsection{Computational aspects}\label{subsec:2.3}
The software presented here is developed under fortran 90 and HPC architecture built with open MPI. For a good performance, the main requirement is to run it on a cluster environment. Next, we describe the main computational features of the code:
\paragraph{HPC architecture} The parallelization is designed differently in the first and second step of the redatuming process. In the first step, each set of receivers is downward continued to the seafloor propagating through the water layer back in time. This is an independent process for each shot gather, and therefore the number of shot gathers are parallelized with the number of available process slots or CPUs at the cluster. On the contrary, in the second step the point gathers are the ones that are independently back propagated through the water layer, and that is why the parallelization is applied in this case to the number of point gathers.

\paragraph{Input/output data} The input field data is directly read in SU format, which is a modification of the standard binary SEG-Y to work with Seismic Unix software \cite{SU}. SU format is similar to SEG-Y but with less amount of information encoded in the headers. In the SU format, each shot consists of a collection of traces, each of them attached to a 240~bytes header encoding experimental information. Apart from reading the traces content, parameters like the shot gather location and the offset at each trace can also be given through the headers. The bathymetry is provided as a text file and the value of the main acquisition parameters (number of receivers, n$_\text{t}$, dt ...) is read also as a text file. Regarding the shot gather results, they are stored also in SU format. We refer to the user's guide for more specific details about the input/output files.

\paragraph{Computational resources} The computational time for running DC depends on the number of CPU's available and the size of our field data. A regular seismic data file usually contains thousands of shot gathers, and a bigger number of point gathers (see Eq.~\ref{Eq:PG}). Taking into account the parallel architecture explained above, the most efficient performance occurs when the first step of the DC is parallelized using the same number of CPU's than shot gathers, and for the second step, the same number of CPU's than point gathers. In this situation, the time to compute the two propagations takes less than 1.5 minutes (in average, each shot or point gather acoustic propagation take from 20 to 40 seconds). Consequently, if running the code on a smaller machine/cluster,  the calculation time is proportional to the number of serial propagations made by each CPU. For example, for redatuming 1000 shots gathers (and $\approx$4000 point gathers) with 50 CPU's, each one computes 20 serial propagations in the first step and 80 serial propagations in the second step. Aside from the propagation time, the time cost of reading/writing field data results inexpensive thanks to the usage of binary formatted files. Regarding the memory space/RAM consumed, this code results inexpensive and low demanding. The biggest objects to be allocated  per CPU are the shot-gather and the water model where the shot gather is propagated, and in the most demanding experimental scenario (deep water, long streamers and a narrow grid model) these arrays occupy less than 20~megabytes.

\subsection{Data pre-processing \textcolor{red}{ }}\label{subsec:2.4}
Field data processing prior to redatuming helps maximize the quality of the results. Here we list some useful processing techniques:
\paragraph{Muting or filtering direct waves} This is important to avoid the screening of the first arrivals in the redatumed shot gathers due to the propagation of the direct wave through the water column. It also allows to clean noise that commonly masks the region, between zero time and the first arrivals. The muting of the direct wave can be done using some form of dip filter or simply by surgical mute.
\paragraph{Band-pass filter} A Butterworth type filter can be applied to remove the lowest frequencies $\le$~2~Hz and frequencies $\ge$~80~Hz to improve signal to noise ratio. 
\paragraph{Spherical divergence} Applying spherical divergence on the data to use realistic amplitudes along the offset.
\paragraph{Improve acquisition geometry}  A low fold acquisition geometry produces redatumed results with a poor signal to noise ratio. Interpolating shot gathers to a closer acquisition geometry, that is, decreasing the distance between receivers and also the distance between sources, hence improving the signal to noise ratio in the redatumed results. This is typically necessary for most pre-90’s legacy data.
\paragraph{Shorten record time} To save computational time, it is plausible to shorten the record time of the input data. This applies when only the time of the first arrivals is needed from the whole redatumed shot gather. A good estimation for this cut-off in order to not eliminate essential information is:
\begin{equation}
\text{time}_\text{cut-off}=\text{FA}^\text{offset=SL}_\text{t} + \text{TWT}_\text{water}
\end{equation}
where $\text{FA}^\text{offset=SL}_\text{vGeom2}$ refers to the estimated first arrival at the end of streamer (SL) in the new virtual geometry~2, and $ \text{TWT}_\text{water}$ the 2-way travel time propagating through the water layer with depth, D:
\begin{equation}
\text{FA}^\text{offset=SL}_\text{vGeom2} = \frac{\text{SL}}{V_\text{Earth}}; ~\text{TWT}_\text{water}=\frac{2\cdot\text{D}}{V_\text{water}}
\end{equation}
As an example, if having field data set with SL~=~6 km, D~=~1 km and V$_\text{Earth}$~=~2~km/s, a cut-off of 4.3 seconds can be applied to the streamer data. A typical record time is 8 seconds, therefore to be able to shorten record times by 50\% in this example, directly would reflect in saving computational resources, including not only RAM but also the back-propagation time at both redatuming steps.

\section{Tests and results}\label{sec:3}
\subsection{Overview}\label{subsec:3.1}
\begin{figure}
\noindent
\makebox[\textwidth]{\includegraphics[scale=0.7]{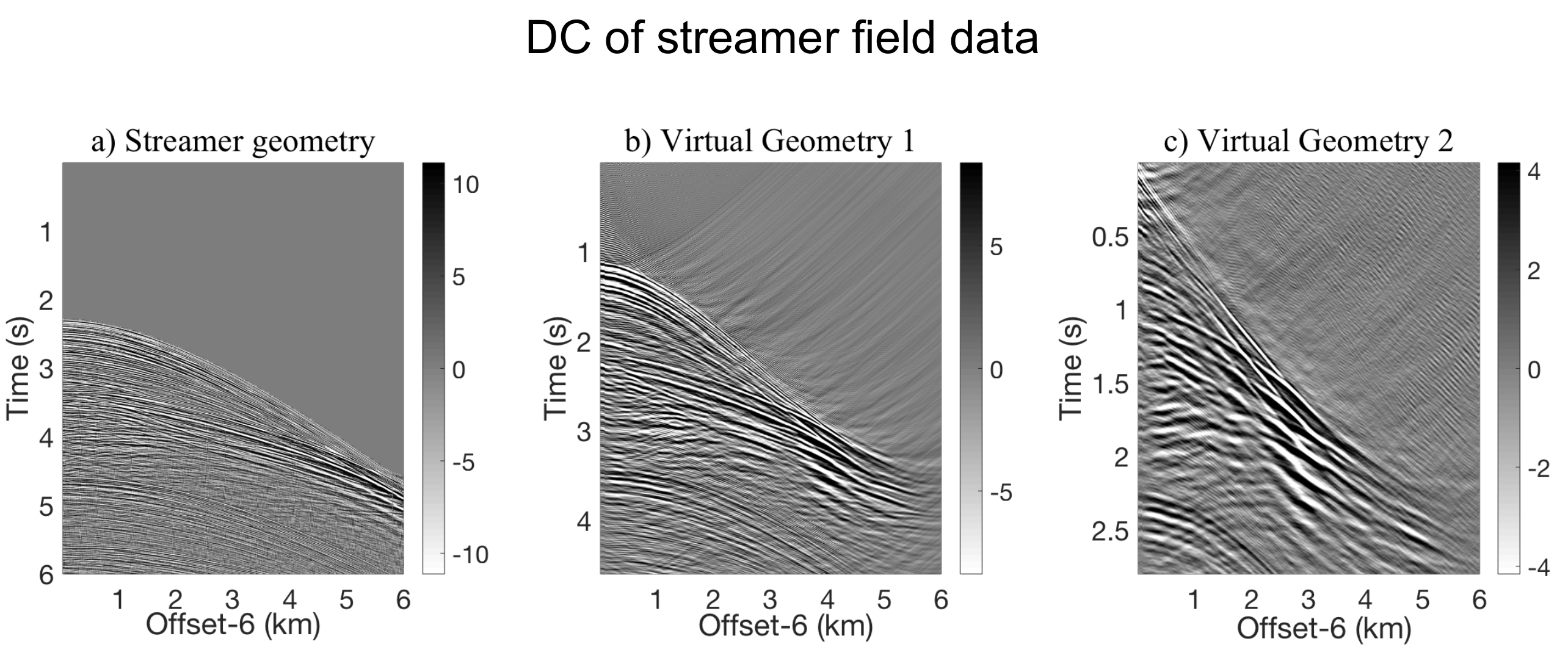}}
\caption[DC overview]{DC redatuming applied to field data recorded on TopoMed experiment~(\cite{topomed_a},~\cite{topomed_b}). For this experiment and location, the streamer length and the water depth are 6 km and 1.6 km, respectively. Panel~a) streamer shot gather. Panel~b) the shot gather in panel~a) redatumed to acquisition geometry~1. Panel~c) the shot gather in panel~a) redatumed to the seafloor, or acquisition geometry~2. }
\label{Fig:intro}
\end{figure}

To introduce the main questions to be addressed in this analysis, we show in Fig~\ref{Fig:intro} the redatuming process applied to a field data case from the TopoMed experiment (\cite{topomed_a}, \cite{topomed_b}). Panel~a) displays the original field data and panels b) and c) show the virtual shot gather redatumed to geometries~1 and~2, respectively. The most remarkable features in Fig.~\ref{Fig:intro} are the full waveform transformation and the water layer decrease of 50\% in the virtual geometry~1 (panel b) and 100\% in the geometry~2 (panel c), respectively. It is also noticeable the short segment of visible refractions in the original data (panel a), around 0.5~km at the end of the streamer. In the virtual shot gather with geometry~2 (panel b), the refractions are visible from 4~km and in the DC shot gather (panel c) the refractions appear as first arrivals from 0~km offset, as expected. Additionally, the appearance of aliasing noise and the odd performance of the redatumed waveforms at long offsets lead to the following questions:
\begin{itemize}
\item Does the total amount of available refractions increase in the virtual redatumed shot gathers?
\item The redatumed data appear to be truncated at long offsets. Why do the phases retract upwards and the amplitudes decrease?
\item Where does the noise in the virtual redatumed shot gathers come from? Can it be reduced?
\end{itemize}
These and more issues are explored in this section. First, we test the performance of the redatuming algorithm with synthetic data. Benchmark models are used to focus on the most relevant attributes of the process and to understand how the refractions and reflections transform at the virtual acquisition geometries~1 and~2. The redatumed results are compared to the ones directly simulated in such configurations. As a ssecond step and to visualize the results on a more realistic geological setting, the redatuming algorithm is applied to data generated using the Marmousi-2 model (\cite{marmousi:2016}). Finally, we show the redatumed results for two different field data sets, TopoMed (already introduced in Fig.~\ref{Fig:intro}) and FRAME experiments, recorded with the same acquisition system and setting but at different water depths.

\subsection{Unraveling the DC redatuming using benchmark models:\newline seafloor refractions and critical offsets} \label{subsec:3.2}
The key point on data redatuming lies on the straightforward visualization of the refractions as first arrivals in the virtual redatumed shot gathers (with geometry 2) from nearly zero offset, in contrast to the substantial amount of refracted waves masked in the original streamer data. Choosing a suitable experimental acquisition configuration is essential to being able to record a significant amount of refractions in the streamer. The empirical ''rule of thumb'' suggests that the streamer length should be at least the water depth to record refracted arrivals. In fact, an optimal streamer length is key to properly redatuming data to the seafloor, as we show below.

In addition to streamer length, several more parameters influence the length of recorded refractions. These are the Vp model structure, the seafloor depth, and the seafloor slope angle. We explore the parameter space with a set of benchmark models, all of them with a constant water velocity, V$_\text{water}$~=~1.5~km/s. The models were generated varying the following parameters:
\begin{itemize}
\item Water column depth, D, from 0.5 to 6~km.
\item One subsurface layer defined with the P-wave velocity, V$_\text{Earth}$, from 2 to 5~km/s.
\item Slope angle, $\phi$, from 0º to 20º.
\end{itemize}
Simple benchmark models permit to predict the critical offset distances; the minimum offset at which refractions are recorded in the streamer data and also the maximum offset at which the first arrivals are truncated in the virtual redatumed shot gathers. Appendix \ref{App:A} shows the analytical calculation of the minimum offset at which refractions appear earlier than reflections for the streamer geometry and the same in appendix~\ref{App:B} for geometry~1. For a flat seafloor ($\phi$~=~0) the minimum offset with refractions, depending on the acquisition geometry, is:
\begin{eqnarray}
	 \text{Offset}_\text{min}^{\text{streamer}}&=&\frac{2\cdot D \cdot u}{\sqrt{1-u^2}} ,~ \label{Eq:r1}\\
	 \text{Offset}_\text{min}^{\text{Geom1}}&=&\frac{D\cdot u}{\sqrt{1-u^2}} ,~ \label{Eq:r2}\\
	 \text{Offset}_\text{min}^{\text{Geom2}}&=& 0 ~, \label{Eq:r3}
\end{eqnarray}
where u~=~V$_\text{water}$/V$_\text{Earth}$. Given a specific streamer length, SL,  the total offset length containing refractions in the streamer is:
\begin{equation}	\label{Eq:r4}
	\text{Offset}_\text{total}=\text{SL}-\text{Offset}_\text{min}^{\text{streamer}} ~\\.
\end{equation}

By applying the redatuming process, the waveform is transformed accordingly to the new datum surface, containing the existing information in the original streamer shot gather. The total length containing refractions in the streamer shot gathers increases when the shot gathers are directly simulated in acquisition geometries~1 and~2, however it is important to understand in this respect, that this total offset length remains the same when the streamer shot gather is virtually redatumed to these geometries. As an example for the case of the virtual geometry~2, the total offset with refractions does not change it only gets displaced to start from the closest receiver to the source. Taking this into account, it is easy to predict the maximum reliable offset in the virtual redatumed shot gathers:
\begin{eqnarray}	
	\text{Offset}^\text{vGeom1}_\text{max}&=&\text{Offset}^\text{Geom1}_\text{min}+\text{Offset}_\text{total} \label{Eq:r5} ~,\\
	\text{Offset}^\text{vGeom2}_\text{max}&=&\text{Offset}_\text{total} \label{Eq:r6} ~.
\end{eqnarray}

Data, including first arrivals, are truncated beyond this point, leading after to the appearance of diffraction tails which are naturally produced by the inverse wave-equation solver. This maximum offset is indeed, the limit-of-use of the whole virtually redatumed shot gather.

It is worth mentioning that the simulation time must be large enough to ensure the recording of refractions and reflections along the entire streamer cable. This can be accomplished by calculating the minimum time required in the equations \ref{Eq:a1} and \ref{Eq:a2} (appendix~\ref{App:A}) with the offset parameter set to the specific streamer length. Setting a long enough recording time guarantees an optimal efficiency of the experimental setup chosen to study a particular geological setting.

\begin{figure}
\noindent
\makebox[\textwidth]{\includegraphics[scale=0.95]{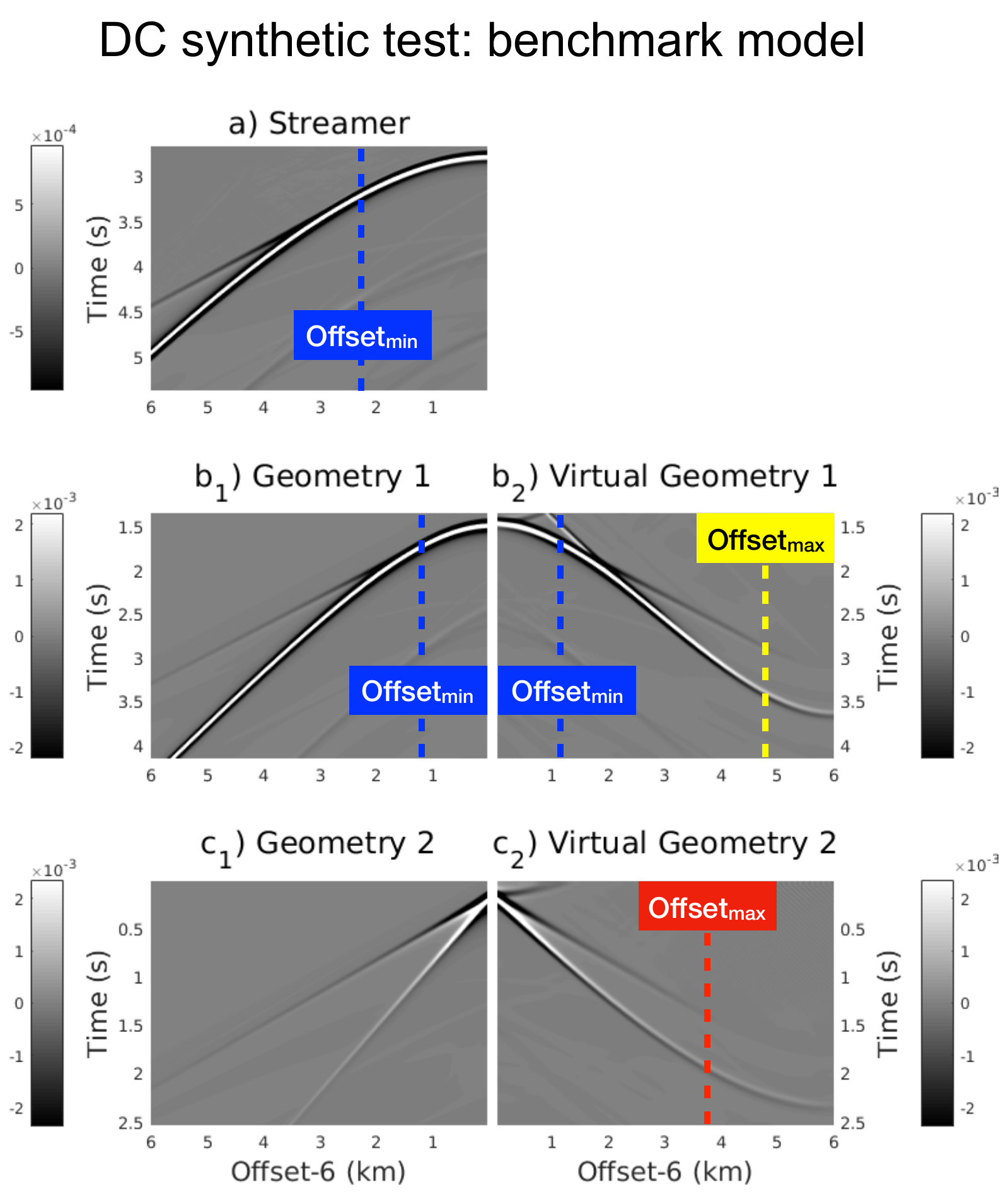}}
\caption[Synthetic test I: one-layer benchmark model]{DC redatuming of a streamer shot gather to the seafloor using a synthetic model with V$_{\text{Earth}}$= 3~km/s and 2~km water depth. The dashed blue line indicate the minimum offset at which early refractions are registered as first arrivals in the streamer (and geometry 1) configuration and the dashed yellow and red lines point at the maximum offset in the virtual redatumed shot gathers (to geometry 2) where the first arrivals are truncated. Following equation 5 in the Supplementary Material, a phase correction of 0.0317 seconds has been applied to the redatumed result in geometry 2.}
\label{Fig:test1}
\end{figure}

As an example, we show in Fig.~\ref{Fig:test1} the DC test applied to one of the benchmark models, with a 2~km seafloor depth, a subsurface velocity of V$_\text{Earth}$~=~3~km/s and a streamer length with SL~=~6~km. Concerning the main acquisition parameters, the simulation time is set to 10~seconds, the source is a Ricker wavelet with central frequency of 10~Hz, the distance between sources is 50~m and between receivers is 25~m. Panels a, b$_1$ and c$_1$ display the shot gathers directly generated in streamer and geometries~1 and~2, respectively. In contrast, panels b$_2$ and c$_2$ display the redatumed shot gather from panel a) into the virtual~1 and~2 geometries. The minimum offset at which refractions are recorded in the streamer geometry is indicated in the figure with a dashed blue line (panel a). In contrast, the dashed yellow and red lines indicate the maximum offset limit for geometries~1 and~2 (panels b$_2$ and c$_2$), respectively. It is apparent the considerable offset length, 1~km, where the refractions remain masked in the streamer (panel a). In the following list, we provide assessment tools to understand whether the DC tool is worth to be applied to a specific field data set:
\begin{table}[t!]
\caption[Outline for the critical offset distances for refractions]{Outline for the critical offset distances for refractions in real and virtually redatumed shot gathers for streamer, and geometries~1 and 2, considering a flat seafloor. SL means streamer length and F~=~Offset$_\text{min}^{\text{streamer}}(\phi=0)$ (see Eq.~\ref{Eq:r1}). It is assumed that SL~$>$~F.}
\begin{center}
\begin{tabular}{c|c|c|c}
Shot geometry & Min offset & Max offset & Total offset \\
&&& \\
\hline
&&& \\
Streamer & F & SL & SL - F \\
&&& \\
Geometry 1 & $\frac{\text{F}}{2}$ & SL & SL - $\frac{\text{F}}{2}$ \\
&&& \\
Geometry 2 & $0$ & SL & SL \\
&&& \\
\hline
&&& \\
Virtual Geometry 1& $\frac{\text{F}}{2}$ & SL - $\frac{\text{F}}{2}$ & SL - F \\
&&& \\
Virtual Geometry 2 & $0$ & SL - $\text{F}$ & SL - F \\
\end{tabular}
\end{center}
\label{Tab:2}
\end{table}

\begin{figure}
\noindent
\makebox[\textwidth]{\includegraphics[scale=0.6]{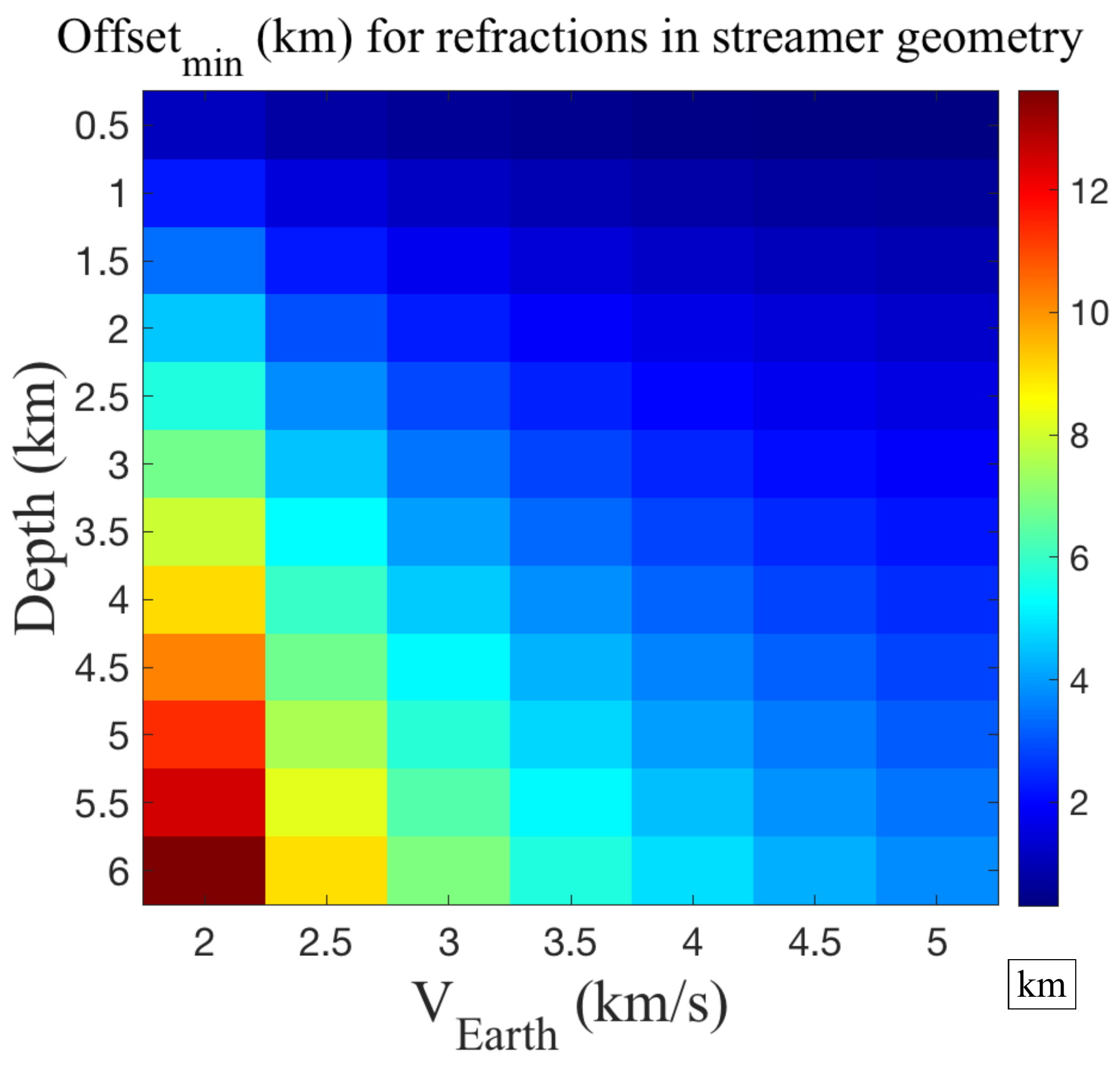}}
\caption[Minimum offset with refractions in streamer data]{Minimum offset (see Eq.~\ref{Eq:r1}), at which the refractions are registered in a shot gather with streamer acquisition geometry, for a flat seafloor ($\phi$~=~0). The x-axis displays the seafloor velocity in the benchmark models from 2 and 5~km/s and the y-axis the water column depth between 0.5 to 6~km.}
\label{Fig:min_offset}
\end{figure}

\begin{figure}
\noindent
\makebox[\textwidth]{\includegraphics[scale=0.95]{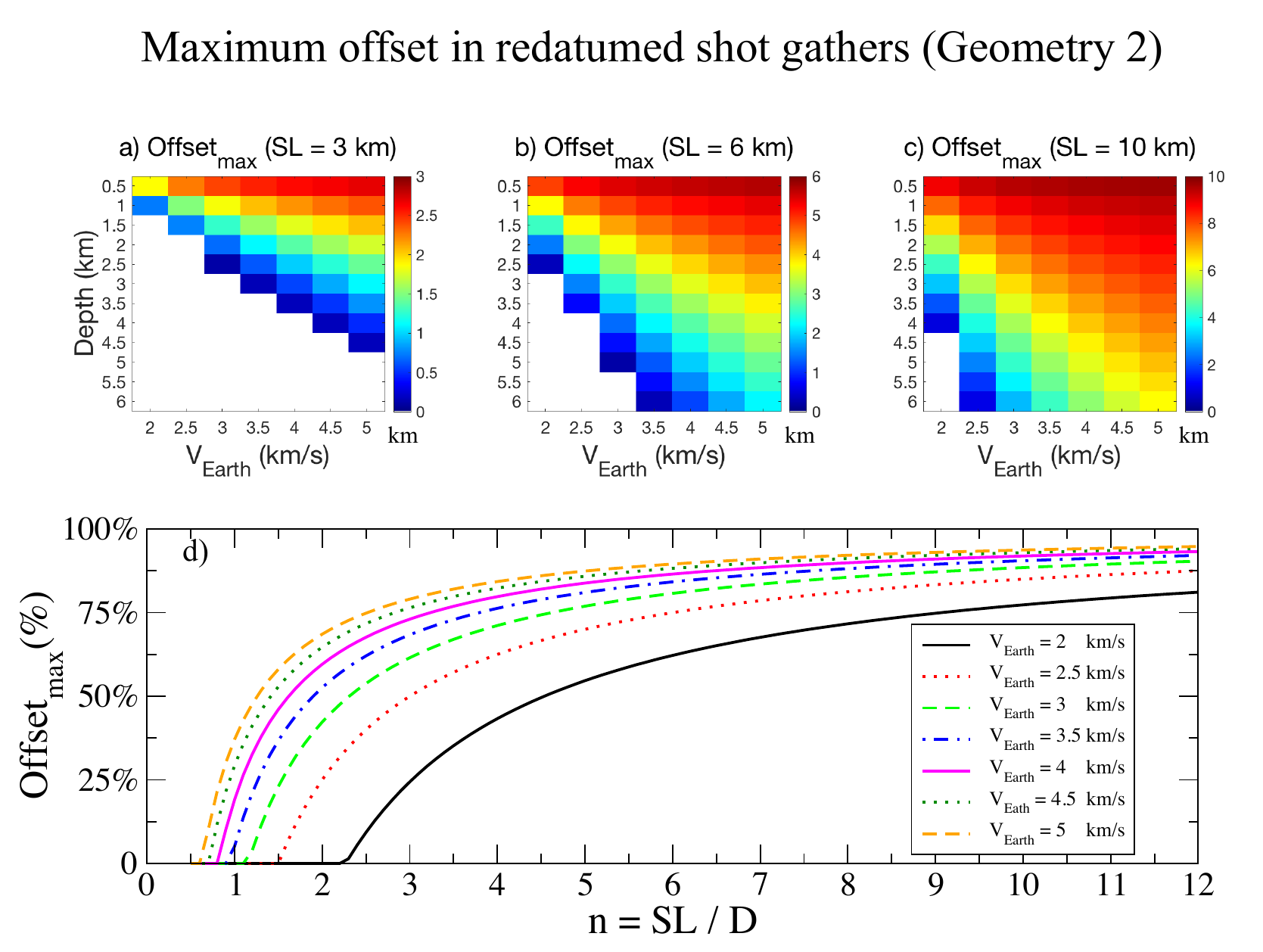}}
\caption[Maximum offset in virtual redatumed shot gathers]{Maximum offset in virtual redatumed shot gathers, Offset$^\text{Geom2}_\text{max}$, where first arrivals are truncated. Upper panels: maximum offset for different streamer lengths (SL) of a) 3~km, b) 6~km and c) 10~km. The x-axis displays the seafloor velocity in the benchmark models from 2 to 5~km/s and the y-axis the water column depth from 0.5 to 6~km. Lower panel d: diagram which summarizes the three upper panels into one. Here the maximum offset is expressed as a percentage (\%) of the total length (y-axis) depending on the rate n~=~$\frac{\text{SL}}{\text{D}}$ (x-axis), which compares the size of the streamer (SL) with the water column depth (D). Every line refers to an specific P-wave velocity of the subsurface, V$_\text{Earth}$.} 
\label{Fig:max_offset}
\end{figure}

\begin{itemize}
\item Tab.~\ref{Tab:2} displays a summary of the critical offset distances and total refraction length in each real and virtual acquisition geometries for flat seafloor marine settings.  
\item The minimum offset at which refractions are recorded in streamer geometry is represented in Fig.~\ref{Fig:min_offset}, for a wide range of depths and subsurface velocities, and for a flat seafloor. The equation~\ref{Eq:r1} is used to compute the minimum offset. This figure helps to understand the challenge and difficulties for recording refractions in marine settings. As expected, there is an inversely proportional decrease of refractions with increasing water depth. Also, the higher the sub-seafloor velocity, the longer the recorded refractions. As an example, the results presented in this figure suggest that in case of having Vp~=~2~km/s and water depth D~=~2~km, the refracted signal cannot be imaged using streamer length $\lesssim$~4.5~km.
\item Upper panels in Fig.~\ref{Fig:max_offset} (a, b and c) shows the maximum offset with refractions in the redatumed shot gathers to virtual geometry~2, Offset$^\text{vGeom2}_\text{max}$ (Eq.~\ref{Eq:r6}). To recreate realistic marine seismic acquisition experiments, we select three different streamer lengths; 3~km (panel a), 6~km (panel b) and 10~km (panel c). For each length, results are shown depending on the different P-wave velocities of the subsurface (x-axis) and seafloor depths (y-axis). The areas in white are the domains in which the DC method cannot be applied due to the absence of refractions (Offset$^\text{vGeom2}_\text{max}$ $\geqslant$ SL). As expected, the longer the streamer, the greater the maximum offset in the redatumed shot gathers
\item Lower panel in Fig~\ref{Fig:max_offset} (d) summarize the three upper panels in one. It shows the same parameter, Offset$^\text{vGeom2}_\text{max}$, but depending on the ratio between the streamer length and the water column depth, n~=~$\frac{\text{SL}}{\text{D}}$, and this is done for the different subsurface velocities. This figure helps to understand the cost/efficiency balance depending on the streamer length in a particular geological setting. Clearly, the slope of the different lines is particularly steep for the first n~=~3~steps. This means that if using an streamer length at least 3~times the seafloor depth, it is relatively easy to register refractions in 60\%-70\% of the total streamer length (depending on V$_\text{Earth}$). However, to go beyond 75\% efficiency, the streamer length must be greater than 3 times the water column depth, an unfeasible geometry for deep water.
\end{itemize}

We refer to Supplementary Material to see more tests using benchmark models. In each test, each specific parameter (water depth, P-wave velocity and streamer length) is modified separately to show its influence in the redatumed results. Also, an analysis is done for a 2-layer Benchmark model to study the behavior of the reflections. Regarding the reflections, their transformation into virtual geometries is shown to be suitable whenever they had previously been recorded in the entire streamer length. Also in the Supplementary Material, a phase discrepancy analysis between the expected and the obtained first arrival in the redatumed shotgathers is presented. It is a systematic error which can be predicted and properly applied to correct the phase in the redatumed results, whenever the subsurface model does not have strong gradient velocity variations.

\subsection{DC redatuming applied to a realistic model}\label{subsec:3.3}
\begin{figure}
\noindent
\makebox[\textwidth]{\includegraphics[scale=0.7]{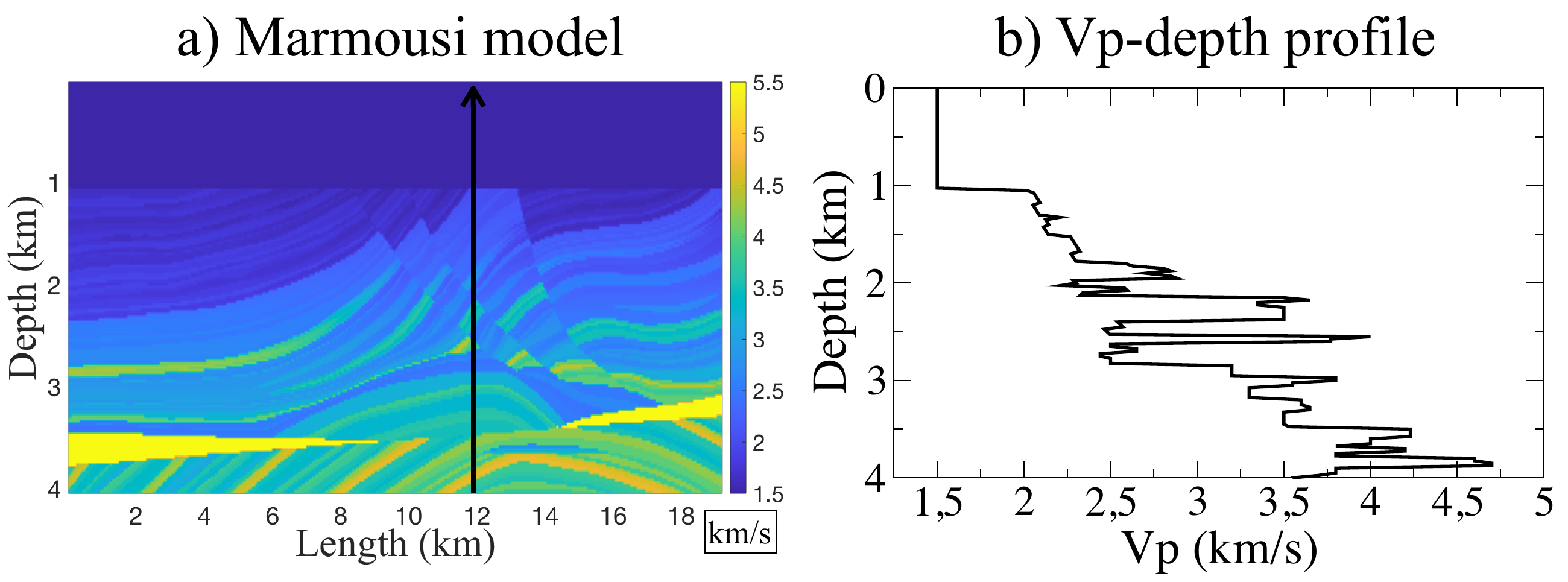}}
\caption[Marmousi model and depth profile]{a) Synthetic Marmousi model with a 1~km water column. The Marmousi model has been re-dimension for this test to the following proportions, 19~km~length~x~4~km~depth. The black line located at around 12~km distance indicates de position at which we show the depth profile in panel b), which is also the position of the shot gather shown in Fig.~\ref{Fig:marm2}. b) Velocity-depth profile at the specific location indicated in panel a) with a black arrow.}
\label{Fig:marm1}
\end{figure}

\begin{figure}
\noindent
\makebox[\textwidth]{\includegraphics[scale=1]{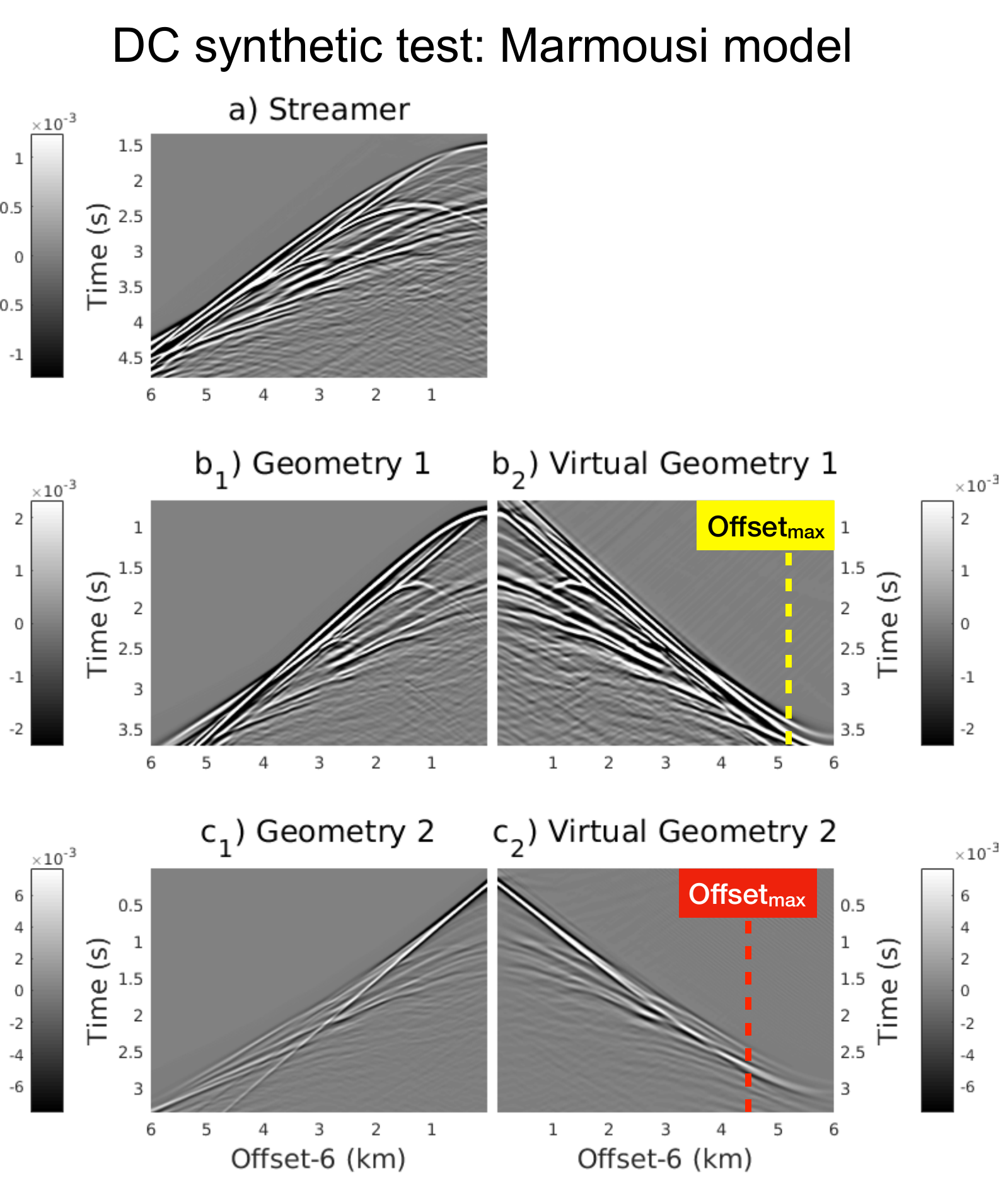}}
\caption[Synthetic test IV: Marmousi]{Redatuming synthetic test using the Marmousi model shown in Fig.~\ref{Fig:marm1}. Panel~a) displays an streamer shot gather generated at the location indicated with a black arrow in Fig.~\ref{Fig:marm1}. Panel~b$_1$ and~b$_2$ show the shot gather generated in geometry~1 and the virtual redatumed result of panel~a) into geometry~1, respectively. Similarly, panels~c$_1$ and c$_2$ show the same as the panels above but for the geometry~2. The yellow dashed line in panel b$_2$) indicate the maximum critical offset for the first arrivals before data are truncated in the virtual geometry~1. The same for panel c$_2$ but for the virtual geometry~2.}
\label{Fig:marm2}
\end{figure}

The sub-seafloor Marmousi model allows to test the redatuming process in a realistic geological scenario. 
For this test, the model has been re-dimensioned to 19~km~length x 4~km~depth and a water column of 1~km has been added (see Fig.~\ref{Fig:marm1}). The simulation time is 10~seconds, the shot point distance 50~m, streamer length of 6~km, receiver distance of 12.5~m, and with a source using a Ricker wavelet centered at 10~Hz. The results are shown in Fig.~\ref{Fig:marm2} for a shot gather placed at 12~km in the model. 

The near offsets record refractions from the shallowest subsurface and the far offsets from deeper parts of the model. As velocity increases with depth, the average velocity of first arrivals increases with offset. This agrees with the results in Fig.~\ref{Fig:marm2}-panel c$_1$, where the first arrivals form a line with a slope of 2~km/s at near offsets and increases up to $\sim$2.7~km/s by the end of the streamer. 

Considering the geological complexity of this model, it is not possible to predict the critical offsets analytically. However, this information can be obtained using Tab.~\ref{Tab:2} with SL~=~6~km, D~=~1~km, and the sub-seafloor velocity which ranges between 2 and 3~km/s (see Fig.~\ref{Fig:marm1}-b). This configuration sets the minimum offset to record refractions in streamer geometry between 1.15 and 2.26~km, and therefore, the maximum offset for geometry~2 with information between 3.74 and 4.85~km. Comparing the shot gather in panel c$_2$ with the shot gather generated virtually with geometry~2 in panel c$_1$, we know that Offset$^\text{vGeom2}_\text{max}\sim$~4.5~km, which fits within the range of possible predicted values. This implies that the minimum offset at which seafloor refractions are recorded in the streamer is 1.5~km.  The fact that refractions were totally masked at this offset and beyond (see Fig.~\ref{Fig:marm2}-a), shows the advantage of applying redatuming techniques.

\subsection{DC redatuming applied to field data}\label{subsec:3.4}
\begin{figure}
\noindent
\makebox[\textwidth]{\includegraphics[scale=0.3]{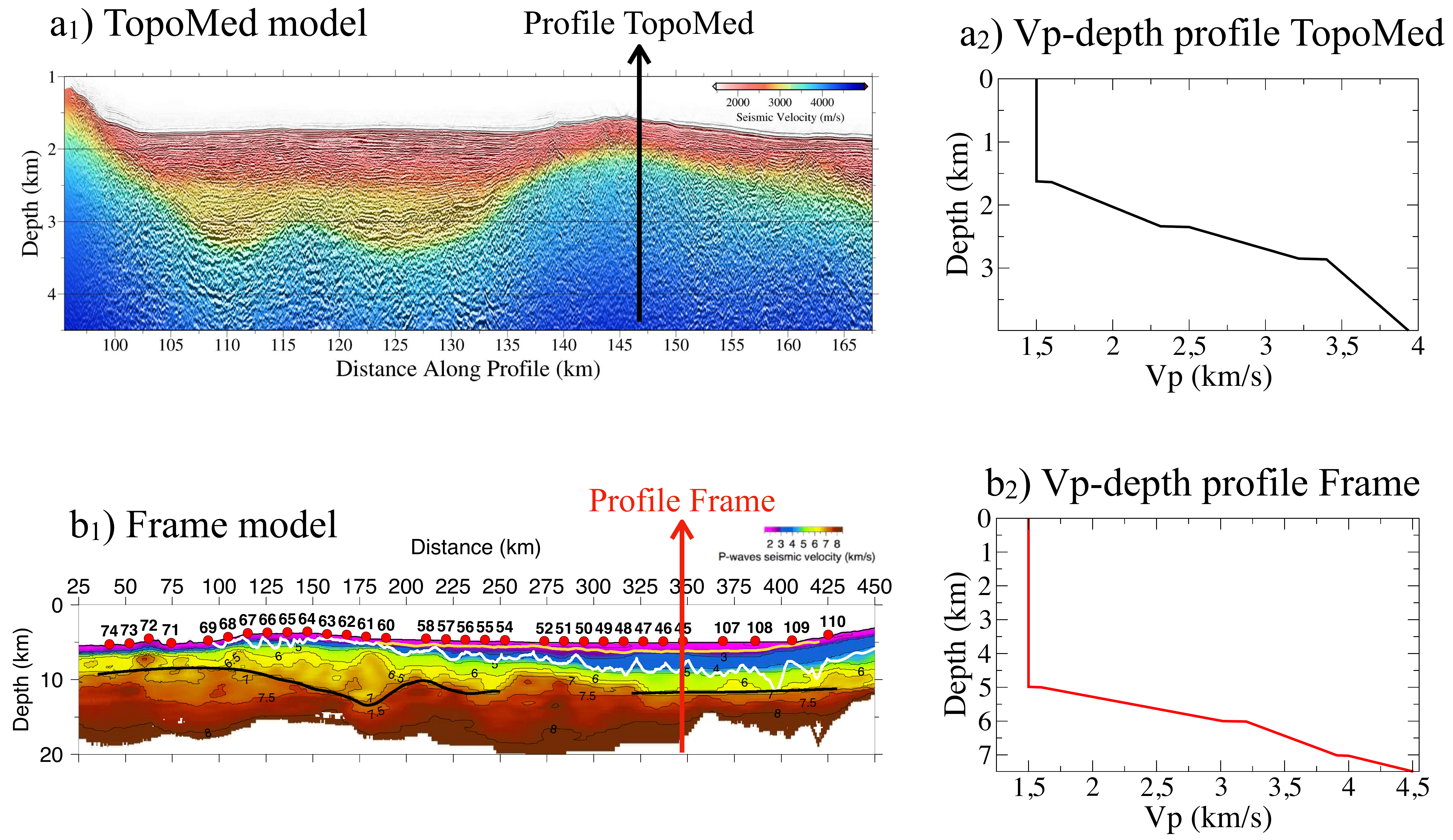}}
\caption[Field data tomographic models]{Field data cases. V$_p$ tomographic models obtained with travel-time tomography for the experiments a$_1$) TopoMed \cite{Gras:2019} and b$_1$) Frame \cite{Merino:2020}. The black and red arrows indicate the location of the depth-profile shown in panels~a$_2$ and b$_2$.}
\label{Fig:field}
\end{figure}

In this section, we apply the DC scheme presented above to field data recorded during FRAME and TopoMed experiments in order to redatuming them to the seafloor. The panels a$_1$) and b$_1$) in Fig.~\ref{Fig:field} display the V$_p$ tomographic models obtained from the seismic records. The black and red arrows indicate de location of the source and depth profiles shown in panels a$_2$) and b$_2$). The TopoMed experiment was conducted in the Alboran Sea (\cite{topomed_a}, \cite{topomed_b}), whereas the FRAME experiment was conducted in the West Iberian margin~\cite{Merino:2021}, and collected by the Spanish R/V Sarmiento de Gamboa. Although the upper kilometres of sub-seafloor velocity gradients of both experiments are similar,(Fig~\ref{Fig:field}, panel c) the water depth of TopoMed profile is ~1.5~km whereas for FRAME it is ~5~km. 

To better understand the redatumed field data results at each experiment, the 1D-depth profiles are used as V$_p$ model to synthetically generate shot gathers with the same acquisition geometry than the original experiment. Then, the redatuming algorithm is applied to the synthetic generated data and also to the field data records. Within each experiment, the results obtained between the redatumed synthetic and field shot gathers are consistent. However, the algorithm performance works very differently for each geological setting.

\begin{table}[t!]
\caption[Parameters for TopoMed and Frame experiments]{Acquisition parameters for the synthetic and field TopoMed and Frame tests. For each case, the specific depth of the field shot gather redatumed, coincident with the position of the 1D-depth profile, is also displayed in the table.}
\begin{center}
\begin{tabular}{c|c|c}
Parameter & TopoMed & Frame \\
\hline
Streamer length & 6~km &~6 km \\
\hline
Distance (sources) & 50~m & 37.5~m \\
\hline
Distance (receivers) & 12.5~m & 12.5~m \\
\hline
Recording time & 8~s & 14~s	\\
\hline
Depth (1D-profile)& 1.62~km & 5~km 
\end{tabular}
\end{center}
\label{Tab:3}
\end{table}

\subsubsection{Well-suited case: TopoMed experiment}
\begin{figure}
\noindent
\makebox[\textwidth]{\includegraphics[scale=0.75]{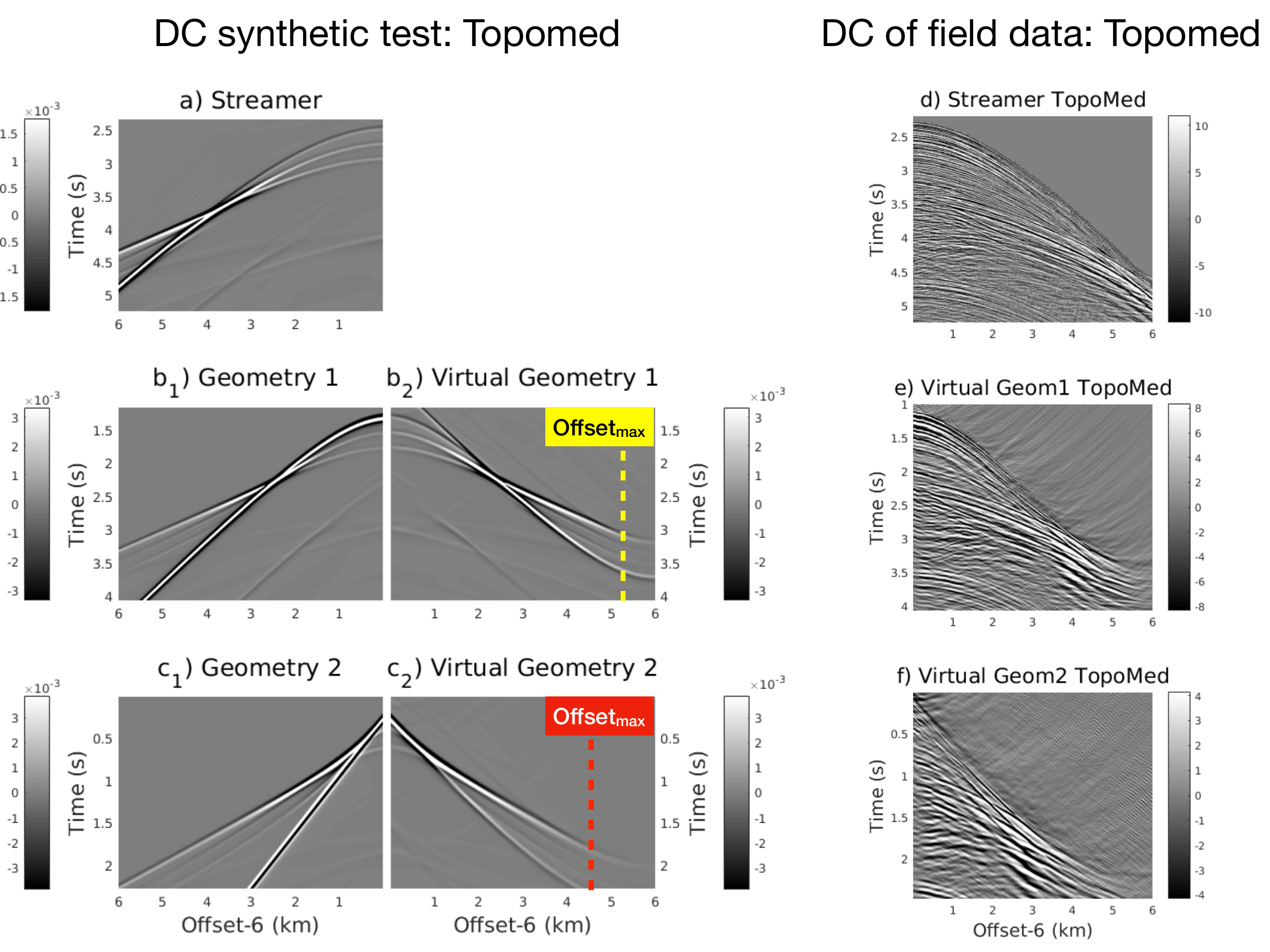}}
\caption[Field data test I: TopoMed]{TopoMed case. Left side: similarly as in Fig.~\ref{Fig:marm2}, it shows the redatuming synthetic test using the 1D depth profile from TopoMed model, shown in Fig.~\ref{Fig:field}. Right side: redatuming of field data from TopoMed experiment; panel~a) field data and panels~b) and c) virtual redatumed results for to the geometries~1 and~2, respectively.}
\label{Fig:field_topo}
\end{figure}

The TopoMed field data records were collected with the acquisition parameters shown in the first column of Tab.~\ref{Tab:3}. The ratio between the streamer length and the water column depth is n~=$\frac{\text{SL}}{\text{D}}$=~3.7, and, as shown in Fig.~\ref{Fig:max_offset}-d), it is an efficient ratio that allows to record refractions in at least 50\%-75\% of the total streamer length. The synthetic test using the TopoMed 1D-depth profile, Fig~\ref{Fig:field_topo} (left panels), shows that the maximum offset in the DC-redatumed shot gather coincides with the range of V$_p$ values provided in Fig.~\ref{Fig:max_offset}, around~4.4~km in panel c$_2$). Similarly for the field data in Fig.~\ref{Fig:field_topo} (panels d-e-f), the virtual redatumed shot gathers transform according to the geometries 1 and 2, showing the critical offset around~4.4~km as well (panel~f).
We clarify that the synthetic and field data cases shown together in Fig.~\ref{Fig:field_topo}, to compare the maximum critical offsets in the virtual geometry 1 results (panels b$_2$ and e) and virtual geometry 2 results (panels c$_2$ and f).

\subsubsection{Poorly-suited case: Frame experiment}
\begin{figure}
\noindent
\makebox[\textwidth]{\includegraphics[scale=0.75]{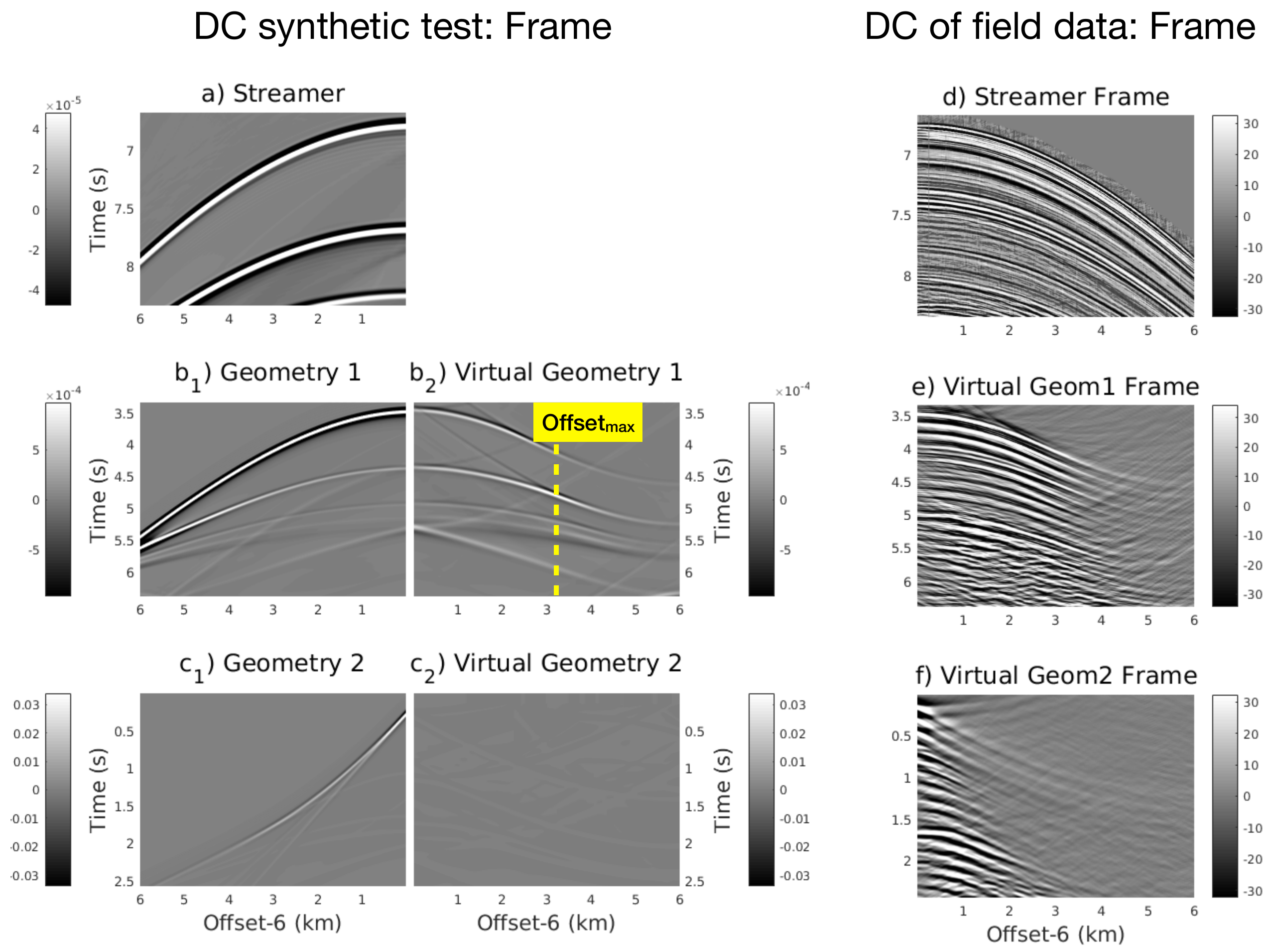}}
\caption[Field data test II: Frame]{Frame case. Left side: similarly as in Figs.~\ref{Fig:marm2} and~\ref{Fig:field_topo}, it shows the redatuming synthetic test using the 1D-depth profile from Frame model, shown in Fig.~\ref{Fig:field}. Right side: redatuming of field data from Frame experiment; panel~a) field data and panels~b) and c) virtual redatumed results for to the geometries~1 and~2, respectively.}
\label{Fig:field_frame}
\end{figure}

Second column of Tab.~\ref{Tab:3} shows the acquisition geometry of the FRAME experiment. In this case, the ratio between the streamer length and the water column depth is n~=~1.2 (see panel d, Fig~\ref{Fig:max_offset}), which indicates a very poor efficiency to register refractions in the streamer and therefore this portends a bad functioning of the redatuming algorithm.
The synthetic test using the model defined with the Frame 1D-depth profile is shown in Fig.~\ref{Fig:field_frame} (left panels). The simulated and redatumed shot gathers in the geometry~1 in panels b$_1$) and b$_2$) are only coincident up to 3~km offset (yellow dashed line). This confirms that the refractions in streamer geometry can only be recorded from 6~km onwards and that the average velocity of the recorded waves is below 3~km/s. This means that there are no refractions originally recorded in the streamer and therefore only reflections are properly redatumed up to the maximum offset in geometry~1. Regarding the geometry~2 in panels c$_1$) and c$_2$), it is apparent that the results only show the propagation tails. Furthermore, the field data redatuming shown in panels e) and f) fails in the same manner too. We can conclude that for marine settings in which the water column depth is 5~km, streamer lengths of at least 10~km should be used (n~=~2) if we aim at recording refractions as first arrivals.

\subsection{Other relevant aspects}\label{subsec:3.5}

\subsubsection{Water column modelling}
\begin{figure}[t]
\noindent
\makebox[\textwidth]{\includegraphics[scale=0.7]{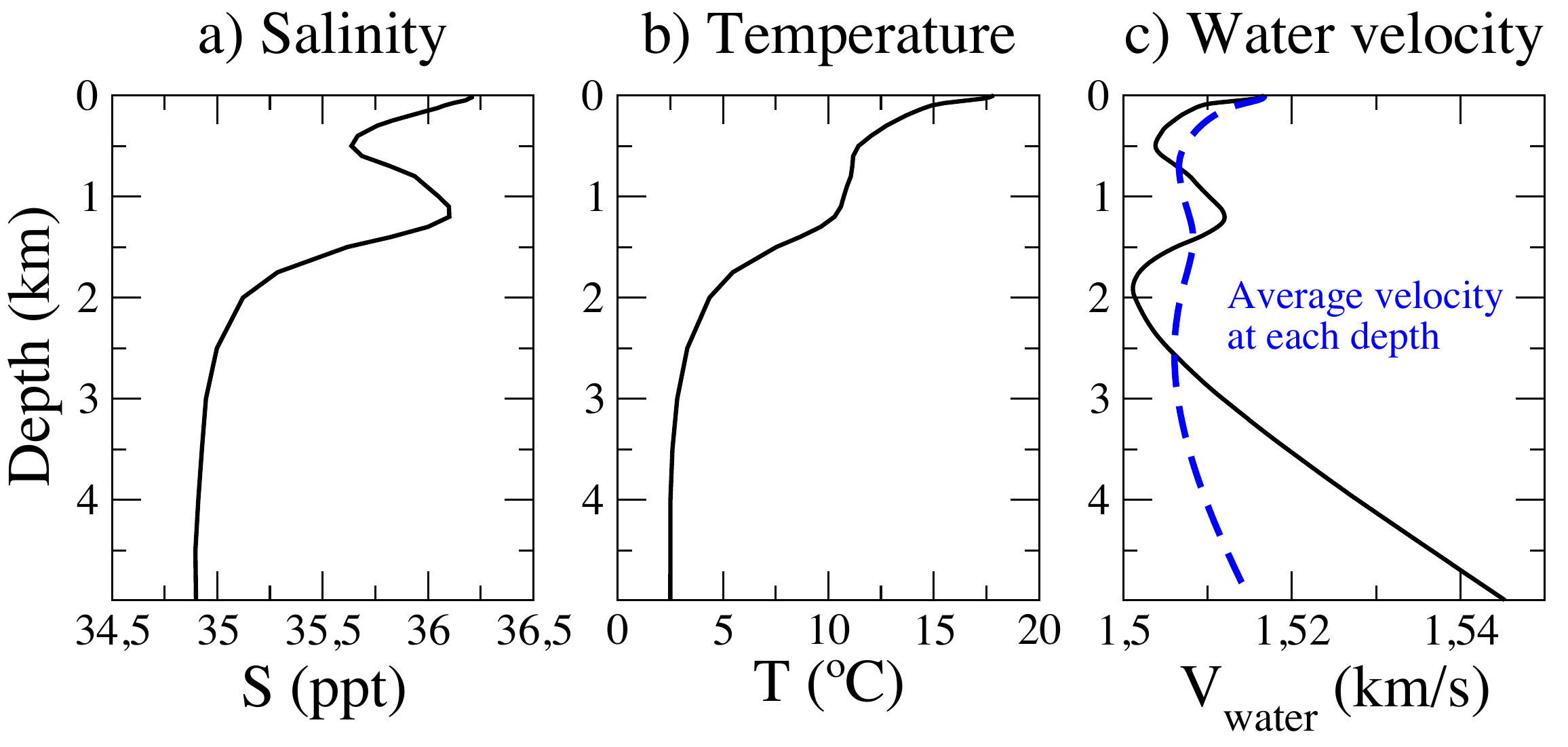}}
\caption[Water column field data: salinity, temperature and velocity]{An example of a) salinity, b) temperature and c) P-wave velocity of the water column. The data for the salinity and temperature are taken from XBT measurements at an specific location in the Atlantic ocean (\href{https://goo.gl/maps/Wt9hC4DCTPHUh4Rt8}{37$^\circ$30'N,12$^\circ$30'W}). The P-wave velocity (black line in panel c) is calculated using the Mackenzie empirical equation~\cite{Mackenzie:1981}. The dashed blue line in panel c) is the average velocity calculated with the velocity values from the sea surface till each specific depth.} 
\label{Fig:waterA}
\end{figure}

For simplicity, in all the tests shown above, the water layer is defined using an homogeneous value for the velocity model V$_{\text{water}}$~=~1500~m/s. However sea water P-wave velocity commonly ranges between 1520~m/s and 1450~m/s. Due to the fact that the DC algorithm presented in this work is based on the 2-way wave equation, it is important to understand the influence of the water layer modelling on the streamer data propagation. As an example, we show in panels a) and b) of Fig.~\ref{Fig:waterA}, XBT data collected for salinity and temperature at an specific point in the Atlantic Ocean (\href{https://goo.gl/maps/Wt9hC4DCTPHUh4Rt8}{37$^\circ$30'N,12$^\circ$30'W}) where the seafloor reaches 5~km depth. The panel c) of the same figure displays the P-wave velocity calculated using the Mackenzie empirical equation~\cite{Mackenzie:1981} (solid black line), which depends on the depth, salinity and temperature. The dashed blue line indicates the averaged velocity from the sea-surface till each specific depth. It is easy to see that this line oscillates around 1511~m/s.

\begin{figure}
\noindent
\makebox[\textwidth]{\includegraphics[scale=0.6]{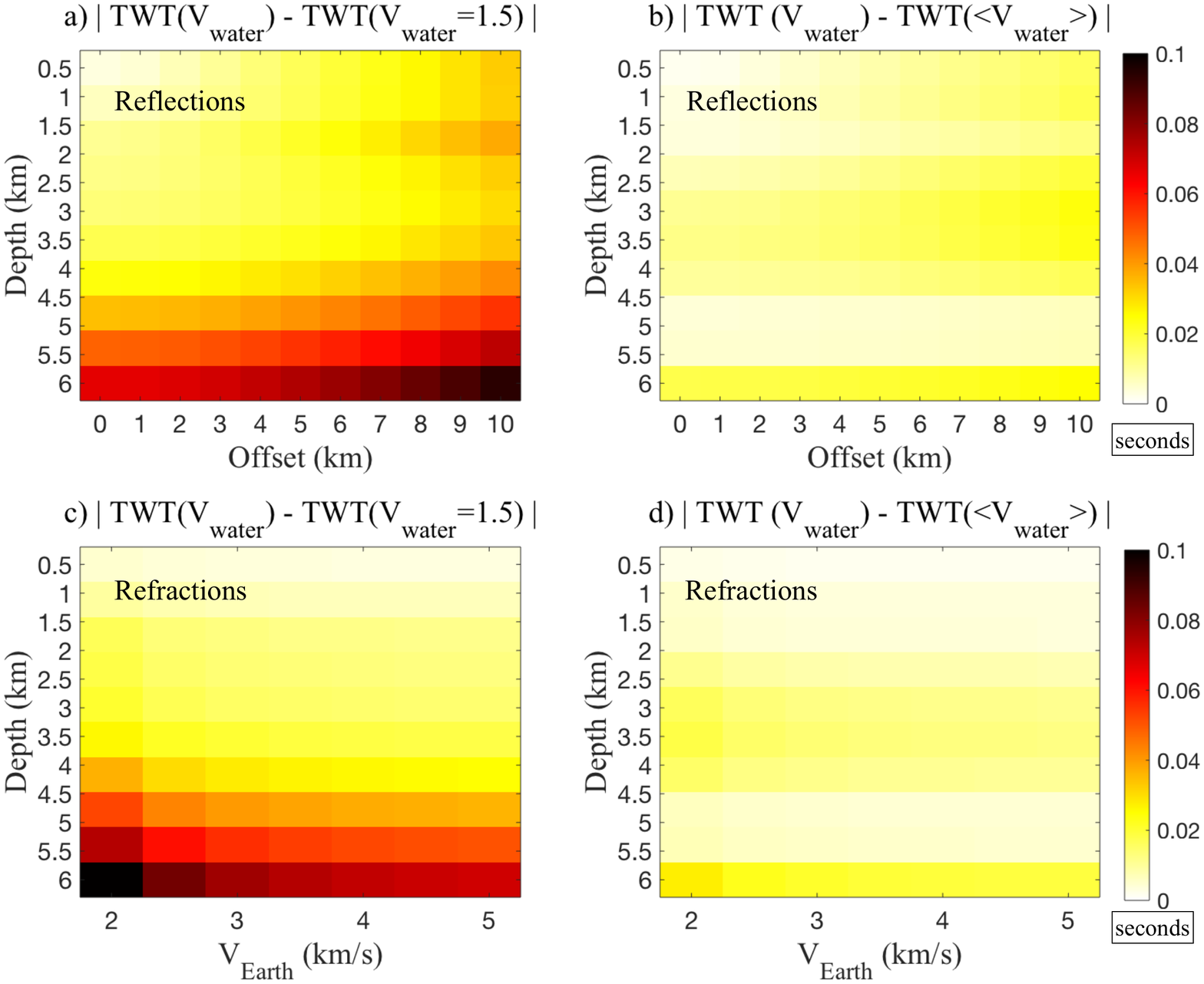}}
\caption[Water column modelling: TWT misfit]{Panels a) and b) show the TWT misfit for seafloor reflections, and panels c) and d) for seafloor refractions, both for $\phi$~=~0. The panels on the left (a and c) compute the misfit between the 'true' water column shown in Fig.~\ref{Fig:waterB} and a homogeneous water column with V$_{\text{water}}$~=~1500~m/s. Similarly in the right panels (b and d) but using for the homogeneous water column the average value of velocity V$_{\text{water}}$~=~1511~m/s.} 
\label{Fig:waterB}
\end{figure}

In Fig.~\ref{Fig:waterB} the TWT misfit is calculated between the true and synthetic water models, for the reflections and refractions from the seafloor. The TWT misfits, taking into account the equations \ref{Eq:a1} for refractions and \ref{Eq:a2} for reflections, are defined as: 

\begin{eqnarray}
\Delta \text{TWT}_{\text{refraction}}&=&| \text{t}^{\text{true}}_{\text{refraction}}-\text{t}^{\text{synth}}_{\text{refraction}} | ,\\
\Delta \text{TWT}_{\text{reflection}}&=&| \text{t}^{\text{true}}_{\text{reflection}}-\text{t}^{\text{synth}}_{\text{reflection}} |
\end{eqnarray}

The propagation in the true water model (black solid line in panel c of Fig.\ref{Fig:waterA}), is compared to two different homogenous synthetic models; the default value in most of the studies, 1500~m/s, and the averaged value for this particular setting, 1511~m/s. The TWT misfit for reflections (panels a and b) is shown for different offsets (x-axis) and seafloor depths (y-axis). The panel a) shows the misfit between the true model and the synthetic one with V$_{\text{water}}$~=~1500~m/s, and similarly in panel b) but for the synthetic model with V$_{\text{water}}$~=~1511~m/s. Also, the TWT misfit for seafloor refractions is shown at the lower panels (c and d), but in this case the x-axis indicates the P-wave velocity of the subsurface, V$_{\text{Earth}}$. The most important observation is that when using a constant velocity model outside the true velocity ranges, the error can reach the order of 0.1~second (2\% in this example). However, if using a suitable P-wave velocity value, the propagation error becomes negligible (0.5\%). Although the algorithm is prepared also to read real data from XBT measurements, for simplicity reasons and on those cases where marine field data is not available, we  recommend to use a realistic constant water velocity value.

\subsubsection{Seafloor slope}
\begin{figure}[t]
\noindent
\makebox[\textwidth]{\includegraphics[scale=0.5]{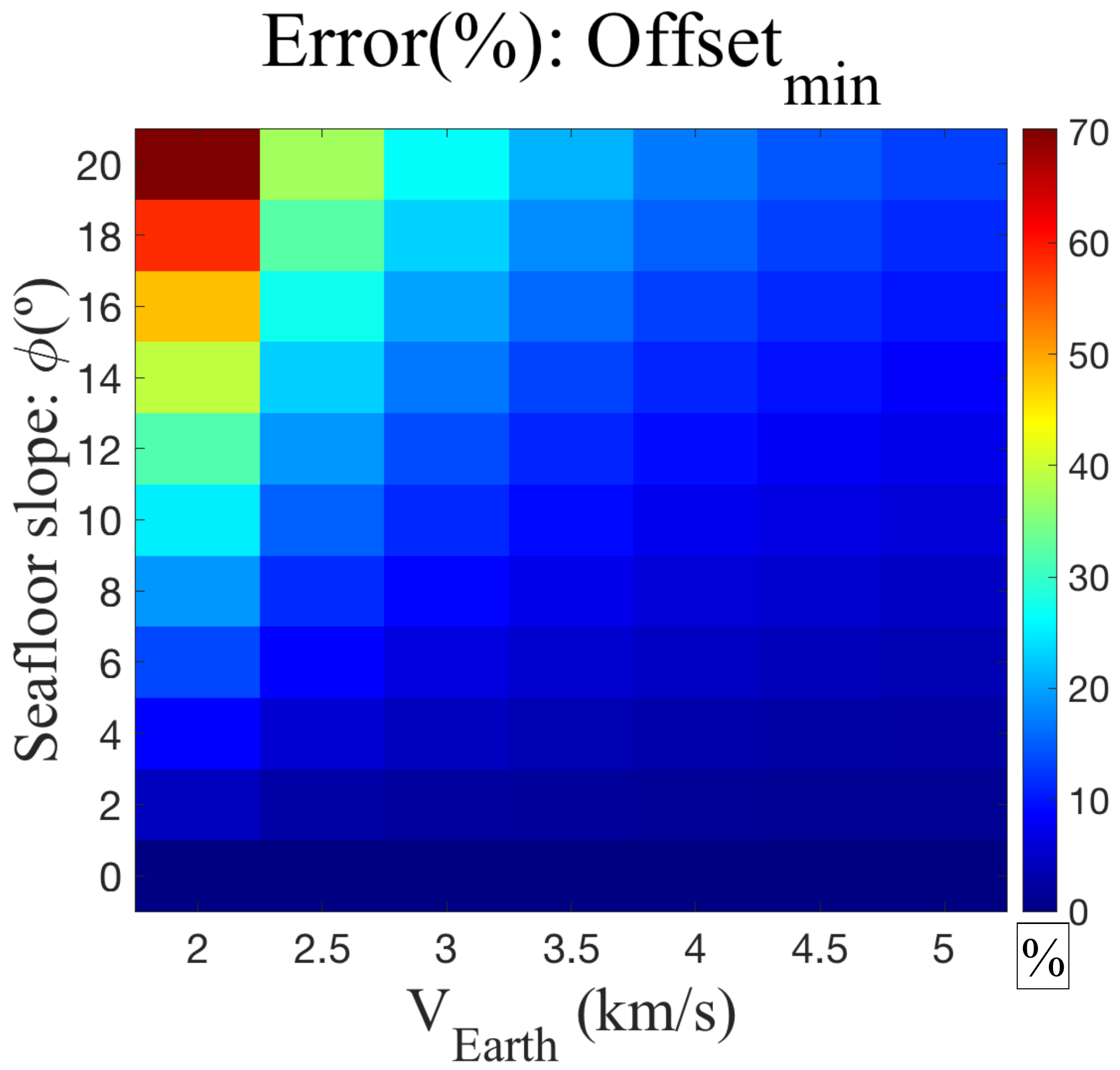}}
\caption[Inclined seafloor: Error(\%)]{Error (\%) of the critical minimum offset to register refractions depending on the P-wave velocity of the subsurface (x-axis) and the angle of the seafloor slope $\phi(º)$ (y-axis). The error formula is computed with eq.~\ref{Eq:}.} 
\label{Fig:tilted}
\end{figure}

The DC algorithm presented can read a specific bathymetry file with any seafloor relief and slope. It is important to note that the shotgather redatuming will be successful only if enough refractions are recorded in the original field data, and this mainly depends on the water column depth, subsurface velocity and streamer length. When the seafloor slope is pronounced, it is convenient to take it into account to properly predict the critical offset. The equation~\ref{Eq:a6} in Appendix~\ref{App:A}, gives the exact prediction for the minimum offset depending on the tilted angle of the seafloor, $\phi$. To understand the effects of the slope of the recorded streamer data, we compute the error formula:
\begin{equation}\label{Eq:}
\text{Error} =\left | \frac{\text{Offset}_\text{min}(\phi)-\text{Offset}_\text{min}(0)}{\text{Offset}_\text{min}(0)} \right | \\.
\end{equation}
This formula is interpreted as the error for the prediction of the minimum refraction offset in streamer data, when the seafloor is assumed to be flat and the slope angle ignored. Fig.~\ref{Fig:tilted} shows as an example of the error(\%) depending on the subsurface velocity (x-axis) and the slope angle $\phi$ (y-axis) for an experimental situation in which $D_s<D_r$ (Eq.\ref{Eq:a6}, appendix~\ref{App:A}). The top limit for the tilted angle is set to 20º which is a realistic upper limit for many submarine slopes~\cite{Harders:2011}.
It is easy to see that for angles, $\phi<$~10º, the total segment of the recorded refractions varies less than 15\% with respect to that recorded for flat surfaces. Therefore, the prediction for $\phi=$~0º is a good first approximation to evaluate field data refractions in common marine scenarios.

\subsubsection{Streamer data with poor acquisition geometry}
\begin{figure}
\noindent
\makebox[\textwidth]{\includegraphics[scale=0.2]{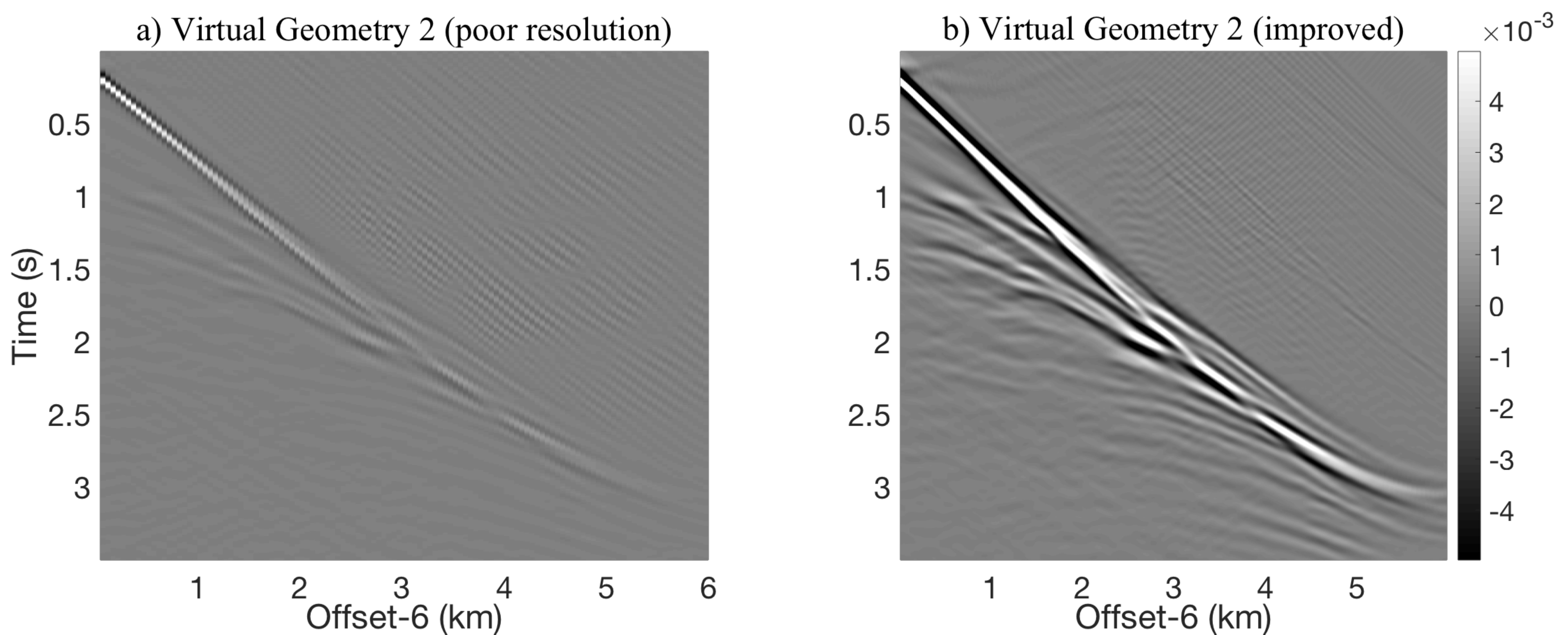}}
\caption[Poor acquisition geometry: legacy data]{Synthetic test simulating legacy field data with a poor streamer acquisition geometry~($d_\text{shots}~=~100~m,~d_\text{rec}~=~50~m$) and using the Marmousi model (Fig.~\ref{Fig:marm1}). Panel~a) shows the virtual DC redatumed shot gather for the poor acquisition geometry, which evidence a very low signal to noise ratio. In panel~b), the signal to noise ratio is clearly improved by interpolating the original streamer shot gathers to a narrower mesh before applying the DC-algorithm as~($d_\text{shots}~=~50~m,~d_\text{rec}~=~12.5~m$).}
\label{Fig:legacy}
\end{figure}

The same test shown for the Marmousi model in Fig.~\ref{Fig:marm2} is repeated here, but to simulate legacy field data, where shot point distance is 100~m and receiver distance 50~m. The challenge for the redatuming algorithm is bigger in this case, as there is less data available due to the sparse acquisition geometry. In Fig.~\ref{Fig:legacy}, we plot only the results obtained after redatuming the shotgather to geometry~2. As expected, the signal to noise ratio is poor (panel a), and aliasing noise is apparent. However, if the original shot gathers are interpolated to halve shot point and receiver distance before the redatuming, the signal to noise ratio improves (panel b). 

\section{Discussion} \label{sec:4}
This work addresses the performance and scope of a redatuming procedure based on the two-way wave equation. The redatuming transforms the streamer field data simulating a virtual seafloor acquisition instead of the real sea surface one, removing the water column from the data. In the particular case of the early refractions, they transform as first arrivals. This provides the advantage to visualize originally hidden recorded refractions, becoming distinguishable as first arrivals in the final virtual configuration. In that respect, the characterization of field data as suitable or unsuitable for redatuming depends on the amount of refractions recorded, as shown in the tests. The recorded length depends on the ratio between the streamer length and the seafloor depth, the subsurface velocity and the seafloor slope.

The quantification of early refractions is central for this analysis due to their importance for building robust velocity models. Besides, it is also the best marker to verify the reliability of the virtual shot gathers and to quantify the maximum offset in all cases. This is possible thanks to the availability of the seafloor bathymetry and the water velocity model, in contrast to the subsurface structure. We have shown that the redatumed results are a space-time transformation of the recorded field data. In this respect, the redatuming does not increase the amount of information at long offsets as if the shot had directly been recorded on the seafloor. For this reason, it is essential to determine as good as possible the maximum reliable offset in the redatumed virtual geometry.

The tests using benchmark models helps to understand that the first point (time and offset distance) where seafloor refractions are recorded in the streamer is the actual starting point of the redatumed shotgather in geometry~2. The first arrivals in this virtual geometry~2 form a line with a slope equal to the value of the velocity of those layers where the refractions travelled, i.e. near offsets register refractions from the shallowest parts of the subsurface and the far offsets from deeper parts. 

Regarding the reflections from the different subsurface layers, the tests with the benchmark model with two subsurface layers (see Supplementary Material) show that the transformation is also suitable up to the maximum offset in virtual geometry~2, whenever the reflection has been previously recorded in the streamer.

With respect to the phase discrepancy of the redatumed results due to the usage of the DC method, is a predictable parameter in geological models with smooth gradient variations, and therefore straightforward to apply to the redatumed results. In the case of realistic scenarios, where the models can be very complex and non smooth, the estimation of phase discrepancies is not so straightforward (see Supplementary Material). Although more tests are needed in this respect, we have observed that this discrepancy is hardly greater than 0.05 seconds.

It is also noticeable the poor transformation of the resultant direct wave traveling through the water layer in the final redatumed shot gathers (see Figs.~\ref{Fig:test1},~\ref{Fig:marm2},~\ref{Fig:field_topo},~\ref{Fig:field_frame} and figures displayed in the Supplementary Material). In this case, the original wave in the streamer data which transforms as the direct wave through the water in the final redatumed shot gather corresponds to waves traveling within the water column. As in the case of refractions, and given a specific geological scenario, this offset increases using longer streamers. However, the offset of the properly transformed direct wave is always shorter than the offset of the properly transformed refractions, and this would be a problem in the case of needing to use the full waveform in the virtual geometry~2.

In this way, although the arrival times in the virtual shot gather and also the reflections are predictable and their reliability easy to quantify, it is not straightforward to understand and control all the details related to the entire waveform. Taking all these aspects into account and the analysis presented in Fig.~\ref{Fig:max_offset}, the best field data candidates to be redatumed to the seafloor are those recorded with streamer lengths two or three times the seafloor depth, and with an average P-wave velocity of the subsurface of V$_\text{Earth}~\geqslant$~3~km/s. Under these conditions, refractions can be recovered in 50\% to 90\% of the streamer length.

The redatuming procedure in this work is based on wave propagation throughout the water column, and in this sense the algorithm is model-dependent. However, as the model refers only to the sea water, where the velocity variation oscillates in a relatively small range of velocity values ($<$ 70 m/s), the absence of detailed information from the water layer is not a drawback to obtain reliable results. If the water column is characterized with a constant but realistic value for the velocity model, the estimated error at a specific offset for reflections and refractions is negligible, from 0.5\%~to~2\%, compared to the value obtained using the real water velocity model.

With respect to noise produced by data aliasing, it is a natural sub-product of the redatuming. The approach to minimise it is data interpolation that increases signal to noise ratio attenuating aliasing noise.

\section{Conclusion}
\begin{itemize}
\item We present and provide an easy-to-use open source software that enables the redatuming of 2D streamer data to the seafloor.
\item  It is a HPC-based code, therefore for an optimal computing time performance it should operate on a cluster.
\item The input field data is required in SU format, which is a common format to store seismic data.
\item Not all field data are suitable to be redatumed, it depends on the amount of refractions originally registered in the streamer. We provide a suit of analytical formulas and assessment visual tools to estimate the suitability of a given data set to perform DC. We show that this depends on the streamer length, seafloor depth, subsurface P-wave velocity and seafloor slope.
\item After applying the DC algorithm the refracted arrivals recorded in the streamer appear as first arrivals from nearly zero offset, so that they are straightforward to pick. This is the main advantage taking into account that a considerable amount of the refractions are screened by reflections and noise in the original streamer recordings. 
\item Apart from the first arrival picking, it is also possible to use the reflection information. However, the waveforms are affected by the redatuming process, so their use is not straightforward.
\item The absence of a detailed velocity model of the water layer is not an impediment to obtain reliable results.  
\item For legacy data recorded with poor acquisition geometry, narrowing the acquisition mesh allows the correct redatuming improving the data contrast and avoiding aliasing noise. This would give a second chance to many past experiments with relevant information in areas where no new data is available.
\end{itemize}

\section{Code availability}
The DC code is available online at GitHub repository: \url{https://github.com/ejimeneztejero/DC}, under the GNU general public license v3.0. 

\section*{Acknowledgments}
This is a contribution of the Barcelona Center for Subsurface Imaging. The work has been partially supported by project FRAME with reference CTM2015-71766-R, funded by the Spanish Ministry of Science, Innovation and Universities. ICM has also had funding support of the ‘Severo Ochoa Centre of Excellence’ accreditation (CEX2019-000928-S), of the Spanish Research Agency (AEI).

\begin{appendices}

\section{Refractions in streamer geometry}\label{App:A}
To calculate the minimum offset in the streamer at which the critical early refractions appear earlier than reflections, one needs to solve the inequality:
\begin{equation}	\label{Eq:a0}
	t_\text{refraction} \leqslant t_\text{reflection} \\,
\end{equation}
where $t_\text{refraction}$ and $t_\text{reflection}$ refers to the TWT from the seafloor, respectively. Following the notation used in the Fig.\ref{Fig:geom}:
\begin{eqnarray}	
	t_\text{refraction}&=& \frac{2\cdot S_1}{V_\text{water}} +\frac{O_1-2\cdot p}{V_\text{Earth}}+ \frac{L_1}{V_\text{water}}, \label{Eq:a1}\\
	t_\text{reflection}&=&\frac{2\cdot S_2}{V_\text{water}} + \frac{L_2}{V_\text{water}} \label{Eq:1b}.\label{Eq:a2}
\end{eqnarray}
Defining $\alpha$ as the incidence angle, $u:sin(\alpha)={V_\text{water}}/{V_\text{Earth}}$, $w:cos(\alpha)=\sqrt{1-u^2}$, and doing some algebra, eq.~\ref{Eq:a0} transforms as:
\begin{equation}	\label{Eq:a3}
	u\cdot O_1 + 2 \cdot d' \cdot w + L_1 \leqslant \sqrt{d'^2+\left(\frac{O_1}{2}\right)^2} + L_2 \\,
\end{equation} 
where
\begin{eqnarray}	\label{Eq:a4}
d'&=&d\cdot\cos\phi ,\\
O_1&=&\frac{ \text{Offset}\cdot \cos\phi \cdot\left( w - u\cdot \tan\phi \right)}{w} ,\\
L_1&=& \frac{ \text{Offset} \cdot \sin\phi}{w} ,\\
O_2&=& \frac{2\cdot d\cdot\cos\phi }{2\cdot d+ \text{Offset}\cdot \tan\phi} ,\\
L_2&=& \text{Offset} \cdot \sin\phi \cdot\sqrt{1+\left( \frac{\text{Offset} } {2\cdot d + \text{Offset} \cdot \tan\phi } \right)^2 } .
\end{eqnarray} 
and $\phi$ is the subfloor slope. Solving eq.\ref{Eq:a3}, the offset minimum at which refractions appear earlier than reflexions is:
\begin{equation}	\label{Eq:a5}
	\text{Offset}_\text{min}^\text{Streamer}=\frac{2\cdot d \cdot u \cdot \left( w - u \cdot \tan\phi \right)}{ u^2 \cdot (\tan^2\phi-1) - 2\cdot u\cdot w \cdot \tan\phi + 1}	\\.
\end{equation}
For any type of marine setting, and eliminating the explicit dependence on D$_r$ as:
\begin{equation}	\label{Eq:a4b}
D_r=D_s+\text{Offset}\cdot\tan\phi\cdot \text{Sign}\cdot[D_r-D_s] \\,
\end{equation} 
The solutions for the minimum offset with refractions, depending on the marine setting type is:
\begin{equation}	\label{Eq:a6}
\text{Offset}_\text{min}^\text{Streamer} = 
\begin{cases} 
D_s \cdot G(\phi) & \mbox{if } D_s<D_r,~(\phi>0)\\
\\
\frac{ D_s \cdot G(\phi)}{1+G(\phi)\cdot \tan\phi} & \mbox{if } D_s>D_r,~(\phi>0)\\
\\
D_s \cdot G(0) & \mbox{if } D_s=D_r,~(\phi=0)
\end{cases}
\end{equation}
where
\begin{equation}	\label{Eq:a7}
G(\phi) = 
\begin{cases} 
\frac{2\cdot u \cdot \left( w - u \cdot \tan\phi \right)}{ u^2 \cdot (\tan^2\phi-1) - 2\cdot u\cdot w \cdot \tan\phi + 1} & \mbox{if } \phi > 0 \\
\\
 \frac{2\cdot u}{w} & \mbox{if } \phi=0 
\end{cases}
\end{equation}

\section{Refractions in geometry~1}\label{App:B}
\begin{figure}
\noindent
\makebox[\textwidth]{\includegraphics[scale=0.65]{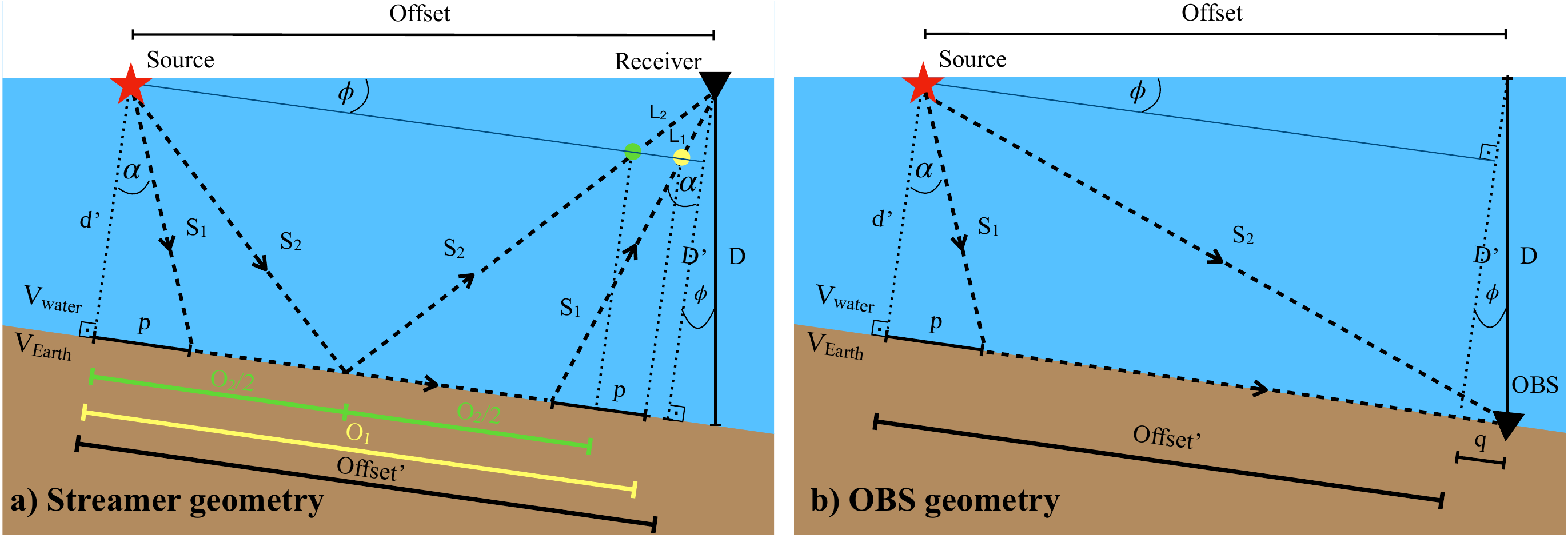}}
\caption[Appendix: refractions in streamer and geometry~1]{Diagrams with the notation used in appendixes A and B for the calculation of the minimum offset with refractions recorded in a) streamer and b) OBS experiments (or geometry 1), respectively. The parameters $V_\text{water}$ and $V_\text{Earth}$ are the water and subsurface P-wave velocities. The angle $\alpha$ refers to the critical incident refraction angle, and $\phi$ to the angle of the seafloor slope. The source is indicated with a red star and the receiver with a black triangle.} 
\label{Fig:geom}
\end{figure}

For the geometry 1 or OBS geometry (sources at the sea surface and receivers at the seafloor), and following the notation of Fig.\ref{Fig:geom}, the TWT refraction and reflection arrivals are:
\begin{eqnarray}	\label{Eq:a8}
	t_\text{refraction}&=& \frac{S_1}{V_\text{water}} +\frac{\text{Offset'}-p+q}{V_\text{Earth}} ,\\
	t_\text{reflection}&=&\frac{S_2}{V_\text{water}}.\\
\end{eqnarray}
Defining $u:sin(\alpha)={V_\text{water}}/{V_\text{Earth}}$ and $w:cos(\alpha)=\sqrt{1-u^2}$, and doing some algebra, the equation to solve, taking into account Eq.~\ref{Eq:a0}, is:
\begin{equation}	\label{Eq:a9}
d' + u \cdot ( \hat{O} \cdot w -u\cdot d') \leqslant w \cdot \sqrt{d'^2+ \hat{O}^2} 
\end{equation} 
where
\begin{eqnarray}	\label{Eq:a10}
d'&=&d\cdot \cos\phi ,\\
 \hat{\text{O}}&=&\text{Offset}\cdot \cos\phi + d\cdot \sin\phi .\\
 \end{eqnarray} 
The solution for the minimum offset with refractions in this specific configuration is:
 \begin{equation}	\label{Eq:a11}
	\text{Offset}_\text{min}^\text{Geom1}= d \cdot \frac{u}{w} - D \tan\phi \\,
\end{equation}
and the general solution at any marine setting, and taking into account Eq.~\ref{Eq:a4b}, is:
\begin{equation}	\label{Eq:a12}
\text{Offset}_\text{min}^\text{Geom1} = 
\begin{cases} 
\frac{ D_s \cdot \left(\frac{u}{w} - \tan\phi \right) }{1+ \tan^2\phi } & \mbox{if } D_s<D_r,~(\phi>0) \\
\\
\frac{ D_s \cdot \left(\frac{u}{w} + \tan\phi \right) }{1+ \tan^2 \phi} & \mbox{if } D_s>D_r,~(\phi>0) \\
\\
D_s \cdot \frac{u}{w} & \mbox{if } D_s=D_r,~(\phi=0) 
\end{cases}
\end{equation}

\end{appendices}

\newpage 

\newpage 

\newpage 

\newpage 

\end{document}


\beginsupplement 


\maketitle

\tableofcontents
\listoffigures

\section{Phase conservation in the redatumed shotgathers}
The purpose of using the DC software is to provide shot gathers collected on a streamer redatumed to the seafloor. In this new datum is easier to visualize and map the recorded refractions as first arrivals in order to invert them with travel-time tomography techniques and therefore obtaining velocity models from the subsurface. In this respect, the concern for the conservation of the full shot gather (amplitudes and phases) in the DC process is reduced to the suitable calculation of the phase, specifically of the first arrival times.

We can define the phase error for the virtually redatumed  shotgathers to geometry~2, as the difference between the expected phases of the shotgather directly generated with sources and receivers at the seafloor ($t_{0}^\text{Geom2}$) and the original MCS shotgather virtually redatumed to the seafloor ($t_{0}^\text{vGeom2}$). Using the DC methodology leads to this type of systematic error for the first arrivals, which mainly depends on the offset and the velocity of the subsurface. This error can be estimated for the first receiver and added after as a constant to correct the phases of the entire shotgather. The following expressions are formulated assuming that source and receivers are located at the same depth position below the sea surface in the MCS experiment, and considering a flat seafloor. The expected first arrival (direct ray) for the trace generated directly at the seafloor can be obtained as:

\begin{equation}	\label{Eq:error1}
	t_{0}^\text{Geom2}=\frac{\text{offset}}{\text{V}_\text{Earth}} \\,
\end{equation}

where the offset parameter refers to the distance between the source and the receiver and ${\text{V}_\text{Earth}}$ to the average velocity of the path through which the direct ray has propagated in the subsurface. The expected first arrival of the original MCS trace virtually redatumed to the seafloor is:

\begin{equation}	\label{Eq:error2}
	t_{0}^\text{vGeom2}=t^\text{streamer}_0-\frac{2\cdot\text{D}}{\text{V}_\text{water}} \\,
\end{equation}

where D is the depth between the original source (or receivers at the streamer) and the seafloor, and $t^\text{streamer}_0$ refers to the first reflected ray from the seafloor, propagating from the source to the receiver:

\begin{equation}	\label{Eq:error3}
	t^\text{streamer}_0=2 \cdot \frac {\sqrt{(\text{offset}/2)^2+ \text{D}^2}}{\text{V}_\text{water}} \\.
\end{equation}

Finally, the phase discrepancy can be estimated as :
\begin{equation}	\label{Eq:error4}
	\text{Error}[t_0^\text{vGeom2}]=  t_{0}^\text{Geom2} - t_{0}^\text{vGeom2} \\.
\end{equation}

\begin{figure}[h!]
\noindent
\makebox[\textwidth]{\includegraphics[scale=0.6]{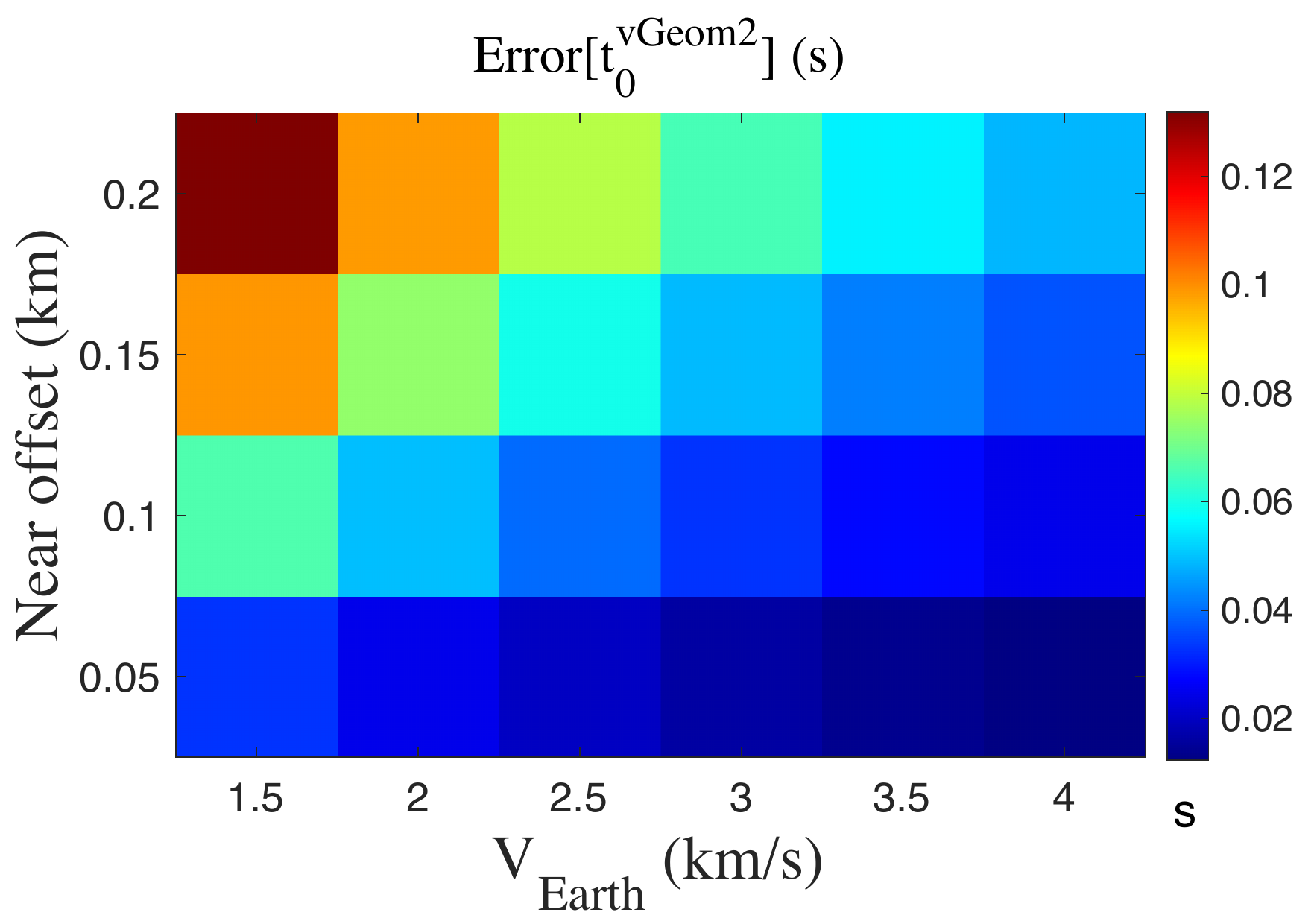}}
\caption[First arrivals discrepancy in redatumed shotgathers]{Time error (in seconds) for the first arrival in the redatumed shotgathers to the seafloor (see equation~\ref{Eq:error4}), depending on the p-wave velocity of the earth surface (x-axis) and the near offset or distance between the source and the first receiver (y-axis). The depth is fixed to 2~km, however the phase error dependency with the depth is almost negligible. }
\label{Fig:error1} 
\end{figure}

This mismatch is plotted in Fig.~\ref{Fig:error1}, depending on the offset and $V_\text{Earth}$ parameters. The influence of the depth parameter in the result is not significant, and for this plot it has been fixed to 2~km. The phase correction to the redatumed shotgathers is also included as an option in the DC software, following the equation \ref{Eq:error4}. 

\begin{figure}
\noindent
\makebox[\textwidth]{\includegraphics[scale=0.5]{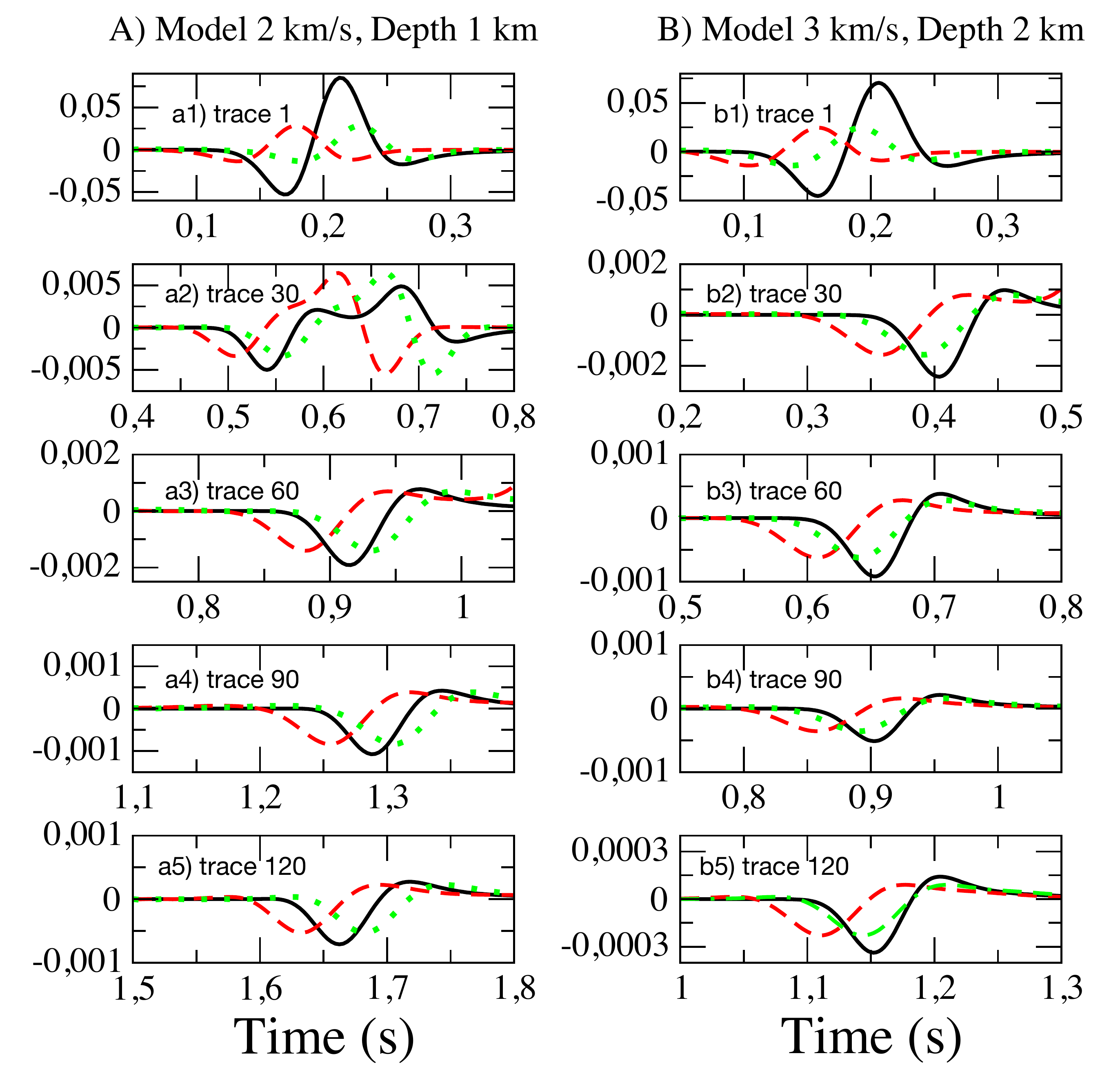}}
\caption[Trace comparison for simple benchmark models]{Trace comparison for two of the benchmark models, both of them using an streamer length of 6 km (240 receivers). Panels in the first column refer to A) benchmark model with V$_\text{Earth}$~=~2~km/s and water depth~=~1~km/s, and those panels in the second column refer to B) benchmark model with V$_\text{Earth}$~=~3~km/s and water depth~=~2~km/s. Black solid lines are the traces synthetically generated directly in Geometry~2, red dashed lines are the traces from the redatumed shotgather to Geometry~2, and green dotted lines are also the traces from the redatumed shotgather but corrected with the error shown in Fig.~\ref{Fig:error1}. Following equation~\ref{Eq:error4}, this correction is 0.0483 seconds for model A) and 0.0317 seconds for model B).}
\label{Fig:error2}
\end{figure}

\begin{figure}
\noindent
\makebox[\textwidth]{\includegraphics[scale=0.5]{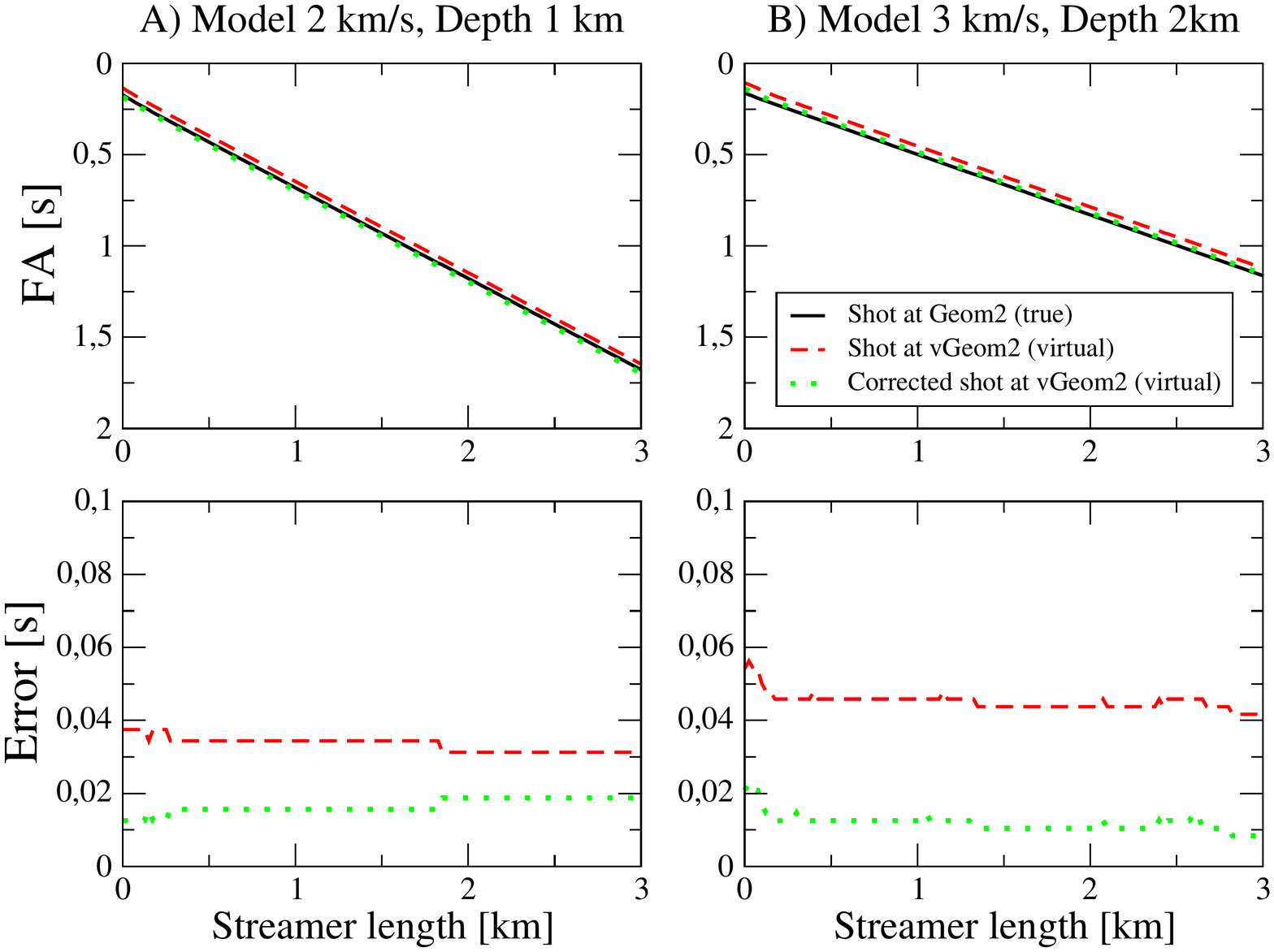}}
\caption[First arrival comparison for simple benchmark models]{First arrival (FA) comparison for two benchmark models. First row shows the first arrivals for three shotgathers; the shotgather directly generated in geometry 2 (black solid line), the virtual redatumed MCS shotgather to geometry 2 (red dashed line) and the redatumed one with the phase correction added (green dotted line). Second row shows the error between the first arrivals of the redatumed shotgathers and the synthetic one. Panels in the first column refer to A) benchmark model with V$_\text{Earth}$~=~2~km/s and water depth~=~1~km/s, and those panels in the second column refer to B) benchmark model with V$_\text{Earth}$~=~3~km/s and water depth~=~2~km/s.}
\label{Fig:error3}
\end{figure}

This phase correction has been successfully applied to the Benchmark tests to better set the phase of the redatumed results to geometry 2. We show in Fig.~\ref{Fig:error2}, for two different benchmark models with the same streamer length (6~km with 240 receivers), different examples of traces from the first receiver to the middle one (1st, 30th, 60th, 90th and 120th). In each panel, three traces are compared; the synthetic trace directly generated at the seafloor (black solid line) and the trace result after applying the DC method (starting from the original trace in the streamer). The raw calculation is shown with a red dashed line, and the result after correcting the phase discrepancy is shown with a green dotted line. In these examples, the phase correction allows to properly set the first arrival to the expected value. In figure \ref{Fig:error3} we show the whole length of first arrival times up to the middle of the streamer (3 km), for both benchmark models (upper panels) and the phase error between the redatumed and expected results (lower panels).

\begin{figure}
\noindent
\makebox[\textwidth]{\includegraphics[scale=0.6]{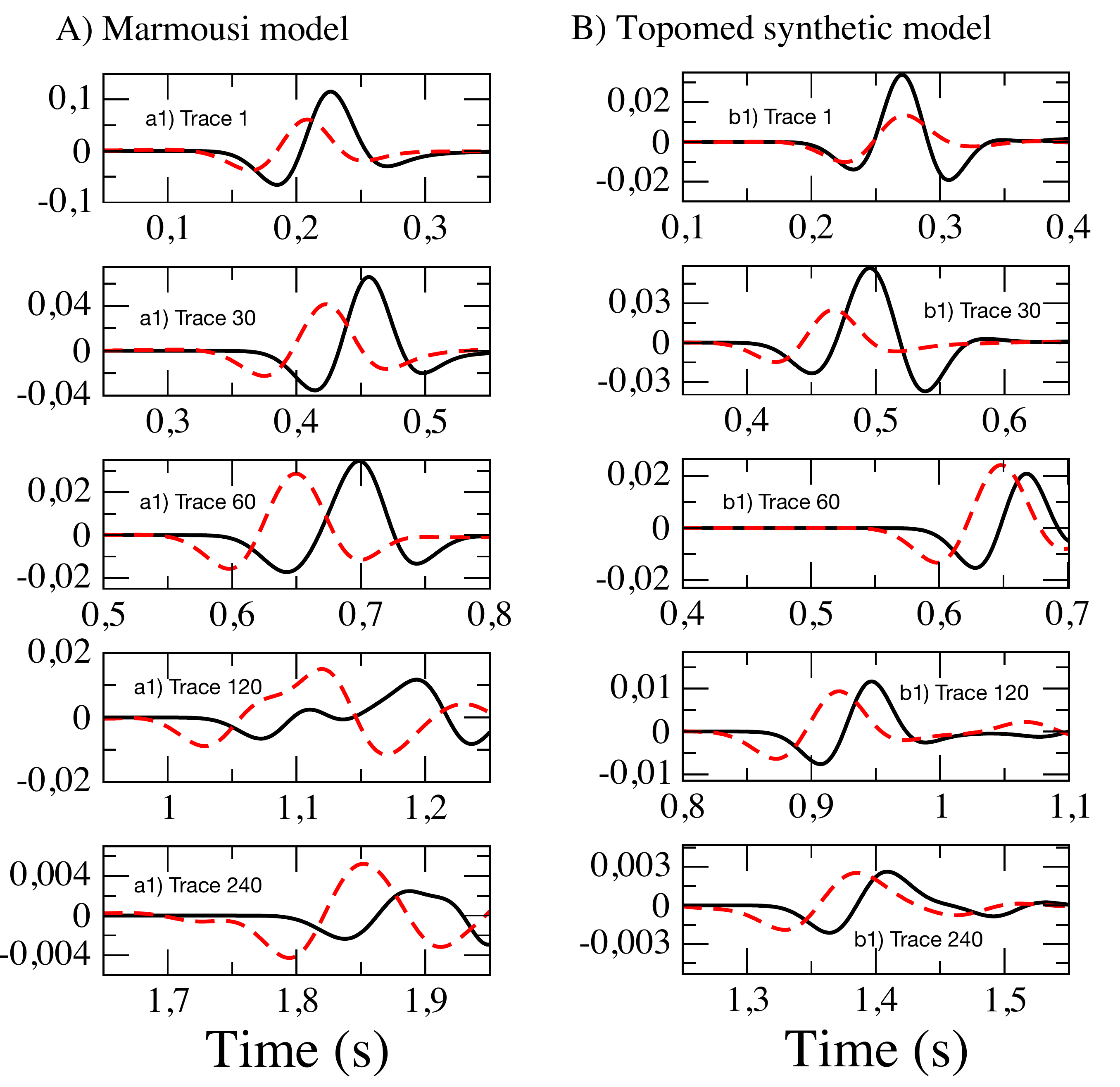}}
\caption[Trace comparison for realistic models]{Similar to Fig.~\ref{Fig:error2} but for different models; panels in the first column refer to A) Marmousi model and panels in the second column refer to B) Topomed synthetic model (see synthetic reference models in the manuscript). In this case, no phase correction has been applied to the redatumed shotgathers.}
\label{Fig:error4}
\end{figure}

\begin{figure}
\noindent
\makebox[\textwidth]{\includegraphics[scale=0.5]{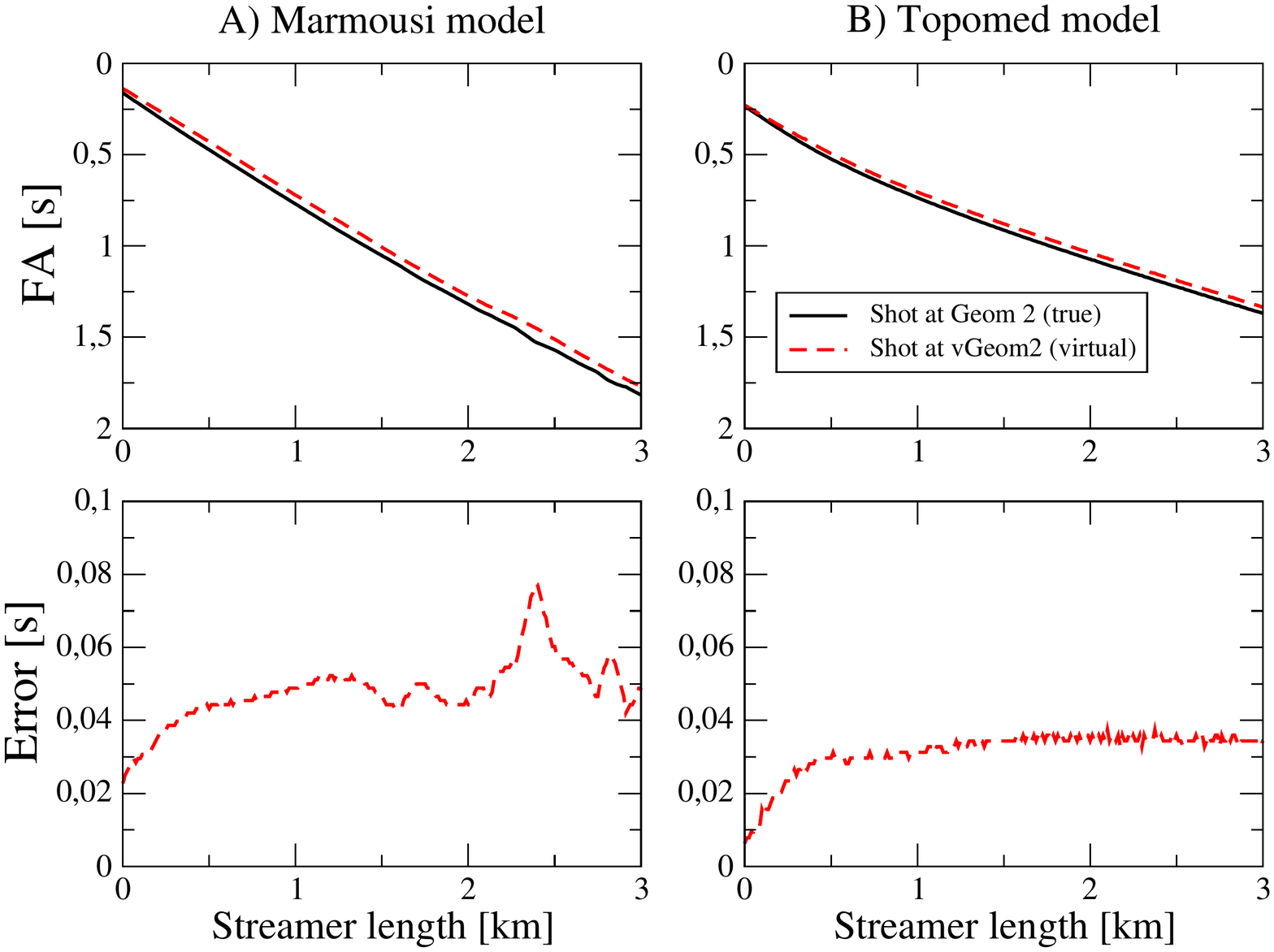}}
\caption[First arrival comparison for simple benchmark models]{Similar to Fig.~\ref{Fig:error3} but for different models; panels in the first column refer to A) Marmousi model and panels in the second column refer to B) Topomed synthetic model (see synthetic reference models in the manuscript). In this case, no phase correction has been applied to the redatumed shotgathers.}
\label{Fig:error5}
\end{figure}

In Figs.~\ref{Fig:error4} and \ref{Fig:error5}, we show a similar comparison but for two other synthetic models; the marmousi model and the Topomed model (see manuscript). In both cases, the streamer length is also 6~km, consisting of 480 receivers. The panels in Figs.~\ref{Fig:error4}  show the comparison between traces from the first to the middle receiver (1st, 30th, 60th, 120th, 240th) and Figs.~\ref{Fig:error5} the length of first arrival times up the middle of the streamer (3 km). 

The main conclusion of the tests is that the phase correction approximation is valid and can be applied in geological scenarios with a smooth velocity gradient. In the case of realistic scenarios, where the models can be very complex and non smooth, the estimation of phase discrepancies is not straightforward. In the examples shown in the manuscript with the marmousi, topomed and frame model, the phase correction is not applied to the redatumed results. However we have observed that this phase discrepancy is hardly greater than 0.05 seconds. Nevertheless, more tests using realistic examples would be needed to analyze and estimate this discrepancy and its possibly predictable behavior, depending on, for example, the velocity gradient of the subsurface. Also, to study the effects that the usage of the virtual redatumed first arrivals with no a priori phase correction would have on the final inverted velocity models after applying tomographic inversion techniques.

\section{DC tests with Benchmark models}
In this analysis, we present the different synthetic tests using the Benchmark models. A total set of 84 models are generated varying the following parameters:
\begin{itemize}
\item Water column depth, D, from 0.5 to 6~km, every 0.5~km.
\item Constant subsurface p-wave velocity, V$_\text{Earth}$, from 2 to 5~km/s, every 0.5~km/s.
\end{itemize}
The common features are:
\begin{itemize}
\item Flat seafloor surface.
\item Constant water p-wave velocity, V$_\text{water}$~=~1.5~km/s.
\end{itemize}

To recreate realistic marine seismic acquisition experiments, we select three different cable lengths for the streamer in our tests; 3, 6 and 10~km. The advantage of using shorter streamers ($\leqslant$~3~km) is that they can be deployed in areas with possible obstacles, providing a good control over the data. Streamers of 6~km length are the most common ones and widely used in academy and industry research. Longer streamers of $\geqslant$~10~km length, mostly used in industry nowadays, provide information from deeper areas of the subsurface but greatly increase the cost and experimental requirements.

Concerning the main acquisition parameters, the simulation time is set to 10~seconds, the source is a ricker wavelet with central frequency of 10~Hz, the distance between sources is 50~m and between receivers is 25~m, and three different streamer lengths are tested: 3, 6 and 10~km. 

The figures~\ref{Fig:test1},~\ref{Fig:test2} and \ref{Fig:test3} show the virtual redatumed shotgathers to geometry 2 depending on the streamer length (SL), the water column depth (D) and seafloor p-wave velocity (V$_\text{Earth}$), respectively. For space and clarity reasons, the results for the real and virtual OBS configurations are omitted in these figures. First and second columns show the shotgather directly simulated with streamer and geometry 2, and the third column the virtually redatumed shotgather from column 1 into geometry 2, respectively. In Fig.~\ref{Fig:test1}, the parameters are set to D~=~2~km and V$_\text{Earth}$~=~3~km/s and SL at each row is 3, 6 and 10~km. As expected, the longer the streamer, the longer the offset with refractions registered. In Fig.~\ref{Fig:test2} the fixed parameters are SL~=~6~km and V$_{\text{Earth}}$~=~3~km/s and D at each row is 1,~2 and 3~km, showing a inversely proportional decrease of refractions with the increasing depth. Finally in Fig.~\ref{Fig:test3}, the fixed parameters are SL~=~6~km and D~=~2~km/s, with V$_{\text{Earth}}$ varying at each row with 2, 3 and 4~km/s. In this case, the higher the velocity of the subsurface, the more refractions registered. 

\section{Two-layer models: reflections and secondary refractions}

Benchmark models with 2-subsurface layers helps to better understand not only the redatuming of reflections but also how the refractions from deeper layers transform into first arrivals. We use a similar synthetic model as the one used in Fig.~2 from the manuscript, but adding a second layer, i.e. a model with water depth of 2~km followed by a first subfloor layer of V$_{1}$~=~3~km/s till 3~km depth and followed by a second layer defined with V$_{2}$~=~5~km/s. The streamer length is 6~km. The redatuming results for the 2-layer model are displayed in Fig.~\ref{Fig:test4}. 

\begin{figure}
\noindent
\makebox[\textwidth]{\includegraphics[scale=0.65]{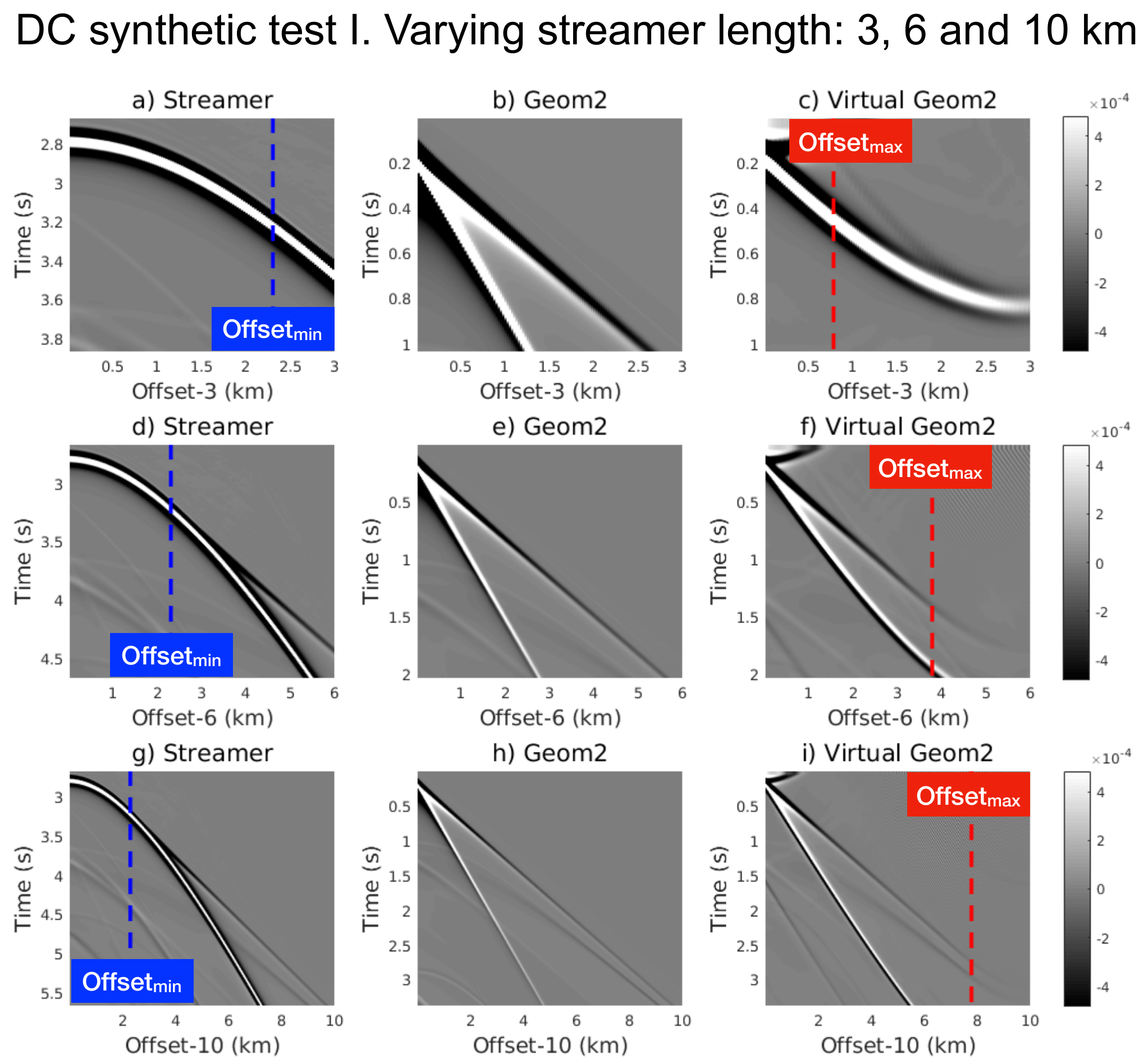}}
\caption[Synthetic test I: streamer length]{DC redatuming of a streamer shotgather to the seafloor (geometry 2) using a synthetic model with V$_{\text{Earth}}$= 3~km/s and 2~km water depth. Every row refers to a different streamer length with the value of~3,~6 and~10~km, respectively. The first and second columns show the simulated shotgather directly with streamer and geometry 2, respectively. The third column shows the results for the virtual redatumed shotgather to geometry 2, after applying the DC-algorithm. The dashed blue line indicate the minimum offset at which early refractions are registered in the streamer configuration (first column, panels a, d, g) and the dashed red line points at the maximum offset in the virtual redatumed shotgather where the first arrivals are truncated (third column, panels c, f, i). The phase correction has been applied to the redatumed results (panels c, f, i), following equation~\ref{Eq:error4}.}
\label{Fig:test1}
\end{figure}

\begin{figure}
\noindent
\makebox[\textwidth]{\includegraphics[scale=0.65]{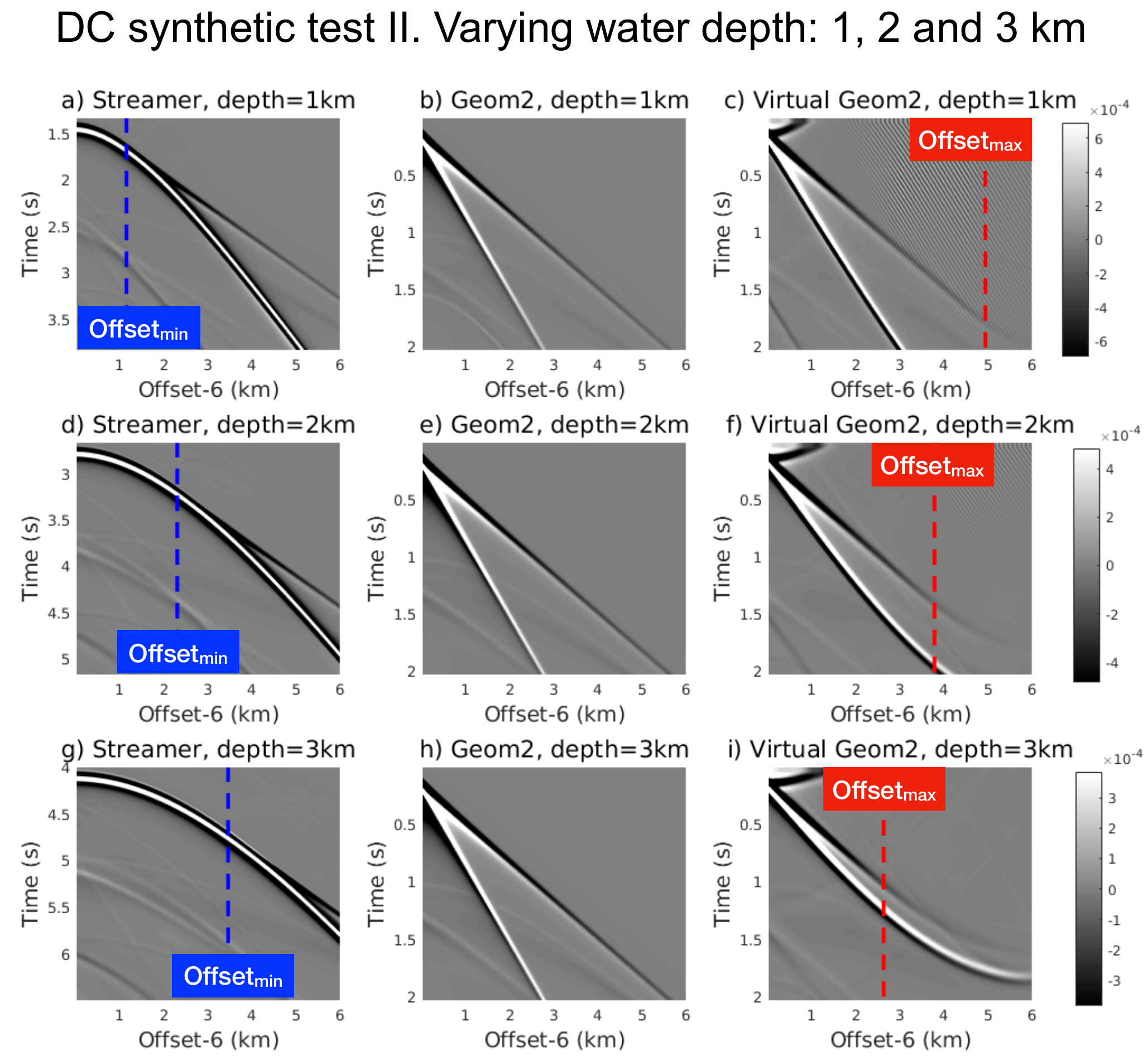}}
\caption[Synthetic test II: water column depth]{Similarly to the test shown in Fig.~\ref{Fig:test1} but in this test, the fixed parameters are V$_{\text{Earth}}$~=~3~km/s and SL~=~6~km. Here, the seafloor depth is the parameter changing at each row with the value of~1,~2 and~3 km, respectively. The phase correction has been applied to the redatumed results (panels c, f, i), following equation~\ref{Eq:error4}.}
\label{Fig:test2}
\end{figure}

\begin{figure}
\noindent
\makebox[\textwidth]{\includegraphics[scale=0.65]{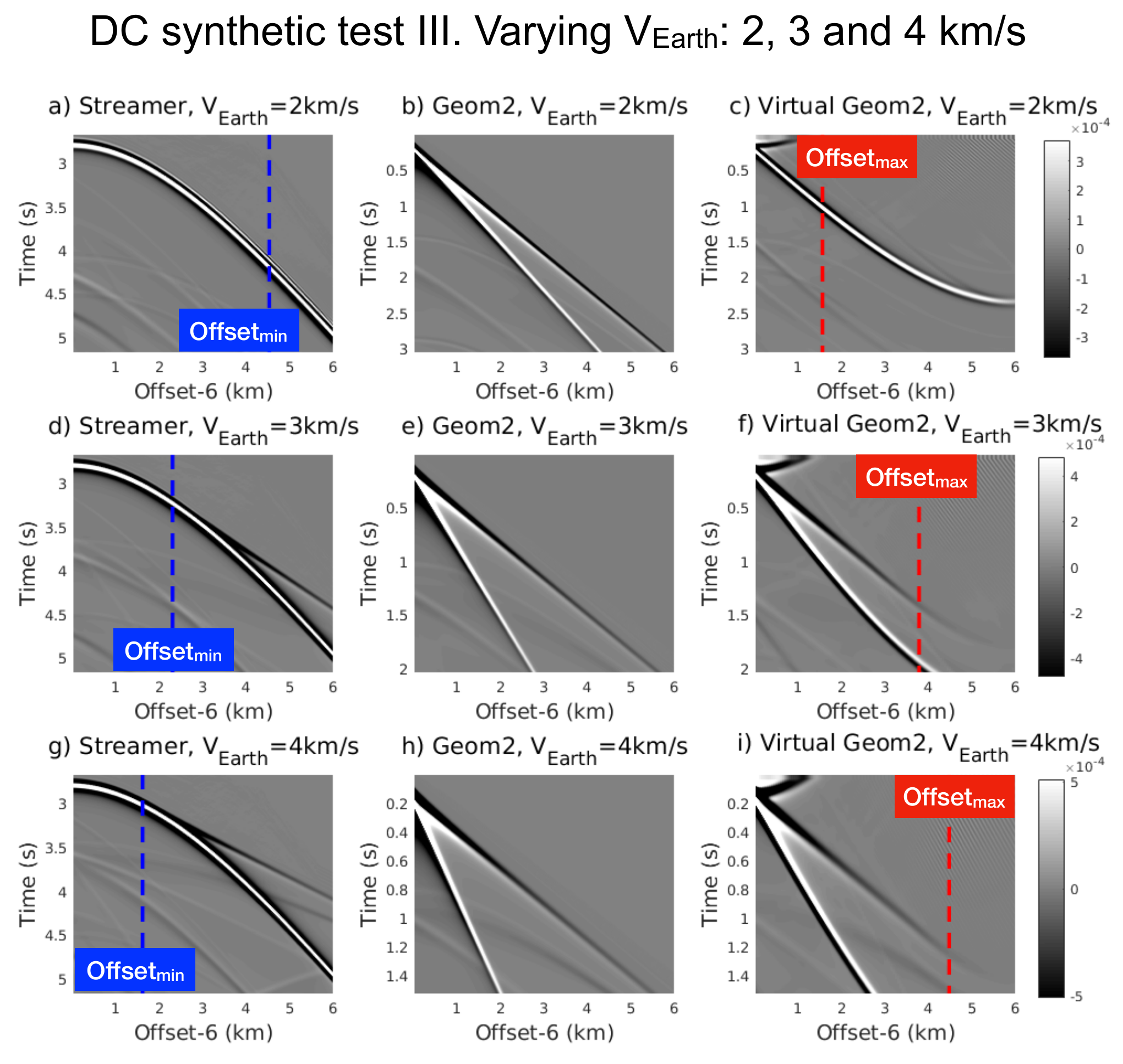}}
\caption[Synthetic test III: subsurface velocity]{Similarly to the test shown in Fig.~\ref{Fig:test1} and~\ref{Fig:test2} but in this test, the fixed parameters are 2~km/s water depth and SL~=~6~km. Here, the subsurface velocity, V$_{\text{Earth}}$ is the parameter changing at each row with the value of~2,~3 and~4~km/s respectively. The phase correction has been applied to the redatumed results (panels c, f, i), following equation~\ref{Eq:error4}.}
\label{Fig:test3}
\end{figure}

\begin{figure}
\noindent
\makebox[\textwidth]{\includegraphics[scale=0.8]{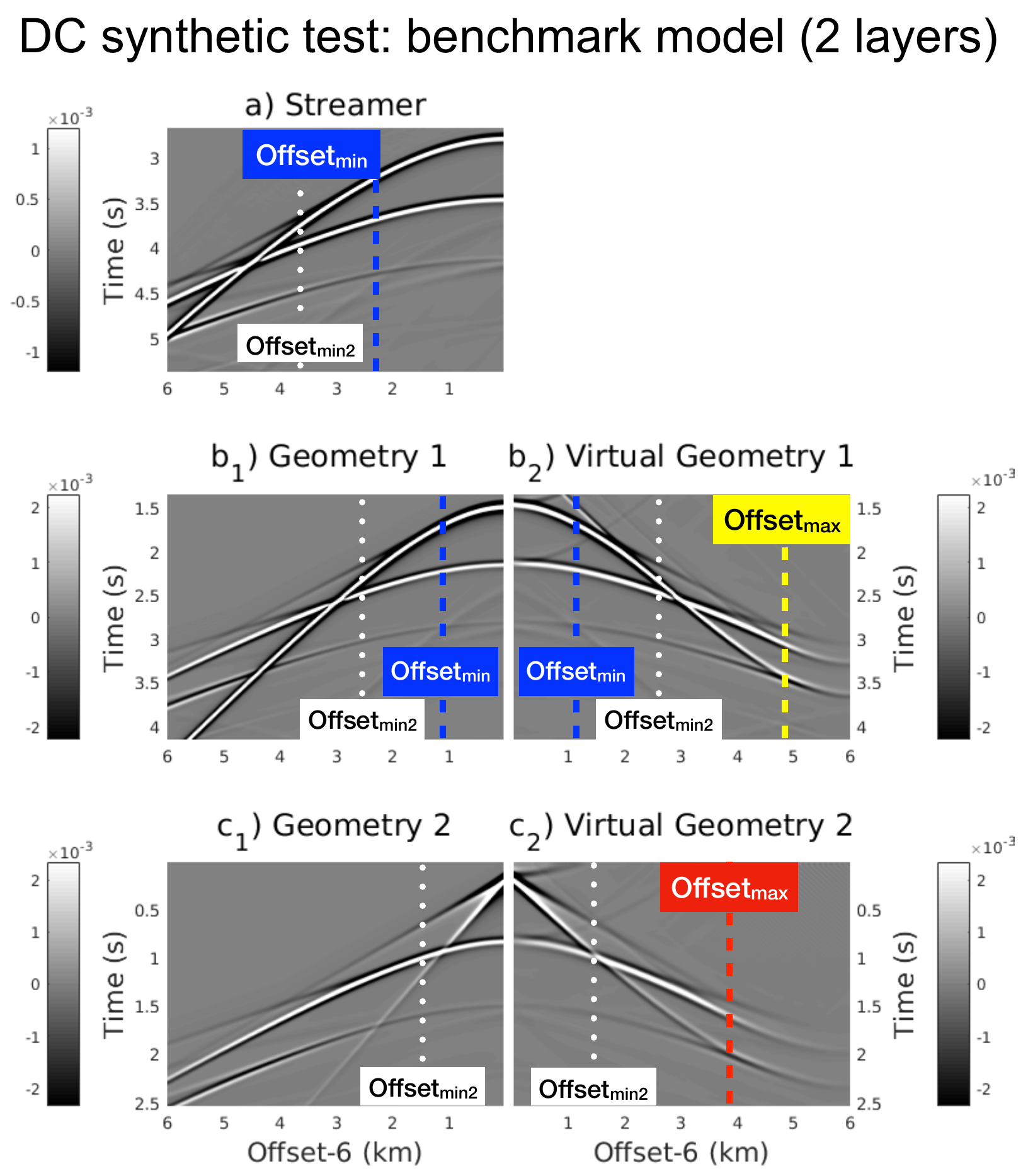}}
\caption[Synthetic test II: two-layers benchmark model]{Similarly to the test shown in Fig.~\ref{Fig:test1} but in this test, there is a second subsurface layer starting at 3~km depth from the sea surface (or 1 km depth from the seafloor) with V$_\text{Earth}$~=~5~km/s. In this case, the added dotted white lines indicate the minimum offset at which refractions from the second layer of the model are registered in the streamer. The phase correction has been applied to the redatumed results (panel c2), following equation~\ref{Eq:error4}.}
\label{Fig:test4}
\end{figure}

To determine the minimum offset for refractions in this configuration, the geometric problem described in the appendix included in the manuscript should be recalculated, but taking into account the refractions coming from the second layer. In any case, the minimum offset will generally be achieved by the seafloor refractions, since to record first the refractions from deeper layers, they should be very close to the seafloor and defined with a higher velocity value compared to the shallow sub-seafloor velocity. This means that also the maximum offset for the virtual geometry 2 is limited in the same way as in the 1-layer type of models (see Figs.~\ref{Fig:test1},~\ref{Fig:test2} and \ref{Fig:test3}).

If the depth of the second subsurface layer is known, it can be calculated the minimum offset where refractions appear earlier than the reflections in the geometry 2, adapting the formula for the minimum refraction offset in streamer geometry:
\begin{equation}
\text{Offset}_{\text{min}_\text{2}}^{\text{DC}}=\frac{2\cdot \hat{D}\cdot u}{\sqrt{1-u^2}} ,~ \label{Eq:1}\\
\end{equation}
where $\hat{D}=\left(D_2-D_1\right)$  is the inter-reflector distance and u~=~V$_\text{1}$/V$_\text{2}$, with V$_\text{1}$ and V$_\text{2}$ the velocities of the first and second subsurface layers, respectively. For the given model parameters the result is $\text{Offset}_{\text{min}_\text{2}}^{\text{DC}}$~=~1.5~km, indicated in the figure with a white dotted line. From this point onwards, the refractions from the second layer might overpass the seafloor refractions and become first arrivals. This can be observed around 3.5~km offset, from where the slope of the first arrivals changes from 3~km/s to 5~km/s. Besides, once $\text{Offset}_{\text{min}_\text{2}}^{\text{DC}}$ is calculated, it is easy to infer it for the rest of the geometries (streamer and OBS). Given that the minimum offset for seafloor refractions in geometry 2 is zero, $\text{Offset}_{\text{min}_\text{2}}^{\text{DC}}$ coincides with the inter-distance between the minimum refraction offsets for the seafloor and for the second layer for any geometry. 

Also, it is noticeable that the maximum offset in the virtual geometries appears extended at least 0.5~km beyond the predicted value. Most of the times, this extra length is picked as real first arrivals, as the phase seems to extrapolate well aligned with the real refractions and also the amplitudes keep their values quite well. This could be considered as a positive secondary effect, depending on the subsurface characteristics. However, the reliable maximum offset should only be extended over the physical limits when the V$_p$ characteristics of the subsurface model are reasonably well known. Picking beyond the predicted offset could produce artifacts in the velocity model, introducing velocities higher than the actual value.

Concerning the reflections, which are always recorded from zero offset, their transformation into virtual geometries is suitable whenever they had previously been recorded in the entire streamer length. In the example of Fig.~\ref{Fig:test4}, the reflections from the second layer are fully recovered up to the maximum offset, Offset$^\text{DC}_\text{max}$. 

